\documentclass[12pt,superscriptaddress,showpacs,showkeys,twocolumn,prb,aps,reprint,a4paper,floatfix]{revtex4-1}

\usepackage{graphicx}
\usepackage{amsmath, amsthm, amssymb, amsfonts}
\usepackage{float}

\usepackage{natbib}
\usepackage{physics}
\usepackage[para,online,flushleft]{threeparttable} 
\usepackage{array}
\usepackage{ragged2e} 

\makeatletter
\newcommand*{\citenst}[2][]{%
  \begingroup
  \let\NAT@mbox=\mbox
  \let\@cite\NAT@citenum
  \let\NAT@space\NAT@spacechar
  \let\NAT@super@kern\relax
  \renewcommand\NAT@open{[}%
  \renewcommand\NAT@close{]}%
  \citet[#1]{#2}%
  \endgroup
}
\makeatother

\makeatletter
\newcommand*{\citenumns}[2][]{%
  \begingroup
  \let\NAT@mbox=\mbox
  \let\@cite\NAT@citenum
  \let\NAT@space\NAT@spacechar
  \let\NAT@super@kern\relax
  \renewcommand\NAT@open{[}
  \renewcommand\NAT@close{]}%
  \cite[#1]{#2}
  \endgroup
}
\makeatother
\usepackage[caption=false]{subfig}

\begin{document}
\title{Exploring the performance of thin-film superconducting multilayers as Kinetic Inductance Detectors for low-frequency
 detection}
\author{Songyuan Zhao}
\email{sz311@cam.ac.uk}
\author{D. J. Goldie}
\author{S. Withington}
\author{C. N. Thomas}
\date{\today}
\affiliation{Cavendish Laboratory, JJ Thomson Avenue, Cambridge CB3 OHE, United Kingdom.}
\begin{abstract}
\noindent We have solved numerically the diffusive Usadel equations that describe the spatially-varying superconducting proximity effect in $\mathrm{Ti-Al}$ thin-film bi- and trilayers with thickness values that are suitable for Kinetic Inductance Detectors (KIDs) to operate as photon detectors with detection thresholds in the frequency range of $50-90\,\,\textrm{GHz}$. Using Nam's extension of the Mattis-Bardeen calculation of the superconductor complex conductivity, we show how to calculate the surface impedance for the spatially varying case, and hence the surface impedance quality factor. In addition, we calculate energy-and spatially-averaged quasiparticle lifetimes at temperatures well-below the transition temperature and compare to calculation in Al. Our results for the pair-breaking threshold demonstrate differences between bilayers and trilayers with the same total film thicknesses. We also predict high quality factors and long multilayer-averaged quasiparticle recombination times compared to thin-film Al. Our calculations give a route for designing KIDs to operate in this scientifically-important frequency regime.
\end{abstract}


\maketitle

\section{Introduction}
\noindent Kinetic Inductance Detectors (KIDs) are thin-film superconducting devices, typically $\sim 100\,\,{\mathrm {nm}}$-thick, that can be configured as ultra-sensitive detectors for astronomical observations across the electromagnetic spectrum. \cite{Jonas_review}
They are readily fabricated by conventional ultra-high vacuum deposition techniques and can be patterned using optical lithography.  KID operating temperatures are low ($T\sim 1\,\,{\rm K}$ or below) giving excellent performance in terms of energy resolution when used as microcalorimeters, or noise equivalent power when used as bolometers. Achieving ultimate performance and low noise requires that the operating temperature satisfies $T/T_{\mathrm{c}}\lesssim 0.15$, where $T_{\mathrm{c}}$ is the superconducting transition temperature. Low operating temperatures ensure sufficiently long quasiparticle recombination times $\tau_{\mathrm{r}}$ such that the quasiparticle generation-recombination noise is minimized and signal-to-noise is maximized.\cite{Leduc_Bumble_TiN,Doyle_2008,Barends_tau_Q_2007,Janssen_2013,deVisser_2014}
 In practice the minimum operating temperature $T_{\mathrm{b}}$ is determined
by available cryogenic technology, so that $T_{\mathrm{b}}\simeq 60\,\,{\mathrm{mK}}$ using an adiabatic demagnetization refrigerator (ADR), or a dilution refrigerator, has become a practical lower-bound for experiments sited in remote high-altitude terrestrial observatories or on satellites.
The photon detection threshold of a KID is determined by the requirement that the photon has sufficient energy to break a Cooper pair $h\nu\ge 2\Delta_{\mathrm{g}}$, where $h$ is Planck's constant, $\nu$ is the frequency of the detected radiation, and $\Delta_{\mathrm{g}}$ is the superconducting energy gap. Each broken pair creates two excess quasiparticles, and the timescale for the recombination of the excess back into pairs, whilst emitting phonons, is determined by
$\tau_{\mathrm{r}}$.
The effect of pair-breaking can be detected by configuring the superconductor into a thin-film $\mathrm{L-C}$ resonator that is readout by a low frequency microwave probe (typically $1-10{\mathrm{\,\,GHz}}$) close to the circuit resonant frequency $f_{0}$.
A change in the number of pairs,
changes the superconductor complex conductivity $\sigma$, surface impedance $Z_{\mathrm{s}}$,
quality factor $Q$, and $f_{0}$, and can be observed in the through-transmission of the probe circuit.
The advantage of the scheme is that as few as two coaxial lines are required to connect from the room temperature to $T_{\mathrm{b}}$ to readout an array. The readout readily lends itself to frequency domain multiplexing.
High detector $Q$ values are required to achieve high responsivity, high sensitivity, and high multiplexing factors.\cite{Jonas_review,Leduc_Bumble_TiN}

There exists a particular challenge in the design of KIDs as low frequency detectors
 for Cosmic Microwave Background observations at
$\sim 70-120\,\,{\mathrm {GHz}}$ from the ground, or on a satellite,\cite{Planck_hifi_2011,Catalano_2015}
for measurement of
low red-shift CO lines at around $100-110\,\,\mathrm{GHz}$,\cite{Cicone_CO_2012,Thomas_2014}
 or for measurements of ${\mathrm O_2}$ rotation lines at $50-60\,\,{\mathrm {GHz}}$ for atmospheric profiling.\cite{Mahfouf_2015,Aires_2015,Turner_2016}
 At the detection threshold frequency $\nu_{\mathrm{g}}=2\Delta_{\mathrm{g}}/h$, the details of the photon-Cooper pair interaction matrix (the so-called case II interaction) mean that the interaction probability is vanishingly-small at low temperatures and increases towards the normal-state value only slowly as a function of $\nu$.\cite{Tinkham_1994}
Using the Bardeen-Cooper-Schrieffer (BCS) result $\Delta_{\mathrm{g}}=1.76\,k_{\mathrm{B}}T_{\mathrm{c}}$ where $k_{\mathrm{B}}$ is Boltzmann's constant, the constraints on $T_{\mathrm{c}}$ determined by bath temperature and pair-breaking threshold can be parameterized as $7\,T_{\mathrm{b}} \lesssim T_{\mathrm{c}} \le \kappa\nu_{\mathrm{g}}$.
$\kappa$ depends on geometry, such that, for direct absorption, $\kappa\simeq 14\,\, \textrm{mK/GHz}$ at threshold, reducing to $\simeq 10\,\, \textrm{mK/GHz}$ if a margin of 25\% is adopted to achieve a pair-breaking efficiency of order 30\% of the normal-state value (see Fig.~\ref{fig:Con} later).
There are few elemental superconductors with the required $T_{\mathrm{c}}$ ($\sim 500-900\,\,{\textrm{mK}}$)
that satisfy these constraints. For example, Al with $T_{\mathrm{c}}$ of $1.2\,\,\textrm{K}$ has a detection \textit{threshold} of $88\,\,\textrm{GHz}$. Current solutions to this problem have been to use alloy superconductors, such as AlMn,\cite{Jones_2015}
 or reactively sputtered materials such as TiN,\cite{Coiffard_2016_TiN}
 such that $T_{\mathrm{c}}$ and $\Delta_{\mathrm{g}}$ can be adjusted to the application. But these sub-stoichiometric compounds have shown wide variations in their properties even in a single deposition, \cite{Vissers_2013} and the current preferred solution is to use multilayers consisting of pure Ti and stoichiometric TiN. \cite{Vissers_highQ_Ti_TiN_2013,Giachero_Ti_TiN}
For transition edge sensors where it is necessary to tune $T_{\mathrm{c}}$, the most-common technological solution is a metallic bi- or trilayer where the properties of a superconducting film ${\mathrm S}$ are modified by the superconducting proximity effect due to its close electronic contact to a lower-gap superconducting or normal metal ${\mathrm S}^\prime$. Cooper pairs diffuse from ${\mathrm S}$ into ${\mathrm S}^\prime$, changing the overall superconducting properties of the composite.
 Catalano~{\textit{et al.}}\cite{Catalano_2015}
 have recently explored this approach in the context of Al/Ti  KIDs. Their prediction for the energy gap of the bilayer used Cooper's model\cite{Cooper_1961} to calculate the weighted-average electron-phonon coupling constant $\overline{N_0V}$ ($N_0$ is the single-spin density of electron states at the Fermi energy and $V$ is the interaction potential). The usual BCS self-consistency expression was then used to solve for $T_{\mathrm{c}}$\cite{Tinkham_1994} and hence $\Delta_{\mathrm{g}}$. We will refer to this approach as the ``weighted-average model''.

A full analysis of a multilayer resonator for use as a KID requires a proper account of the spatial variation of its superconducting properties as a function of position through the film, $x$. The best approach to this problem
is to solve the one-dimensional diffusive
 Usadel equations with appropriate boundary conditions.\cite{Usadel1970,Brammertz2001,Golubov2004,Vasenko2008,Wang_2017}
 This is the approach we use here. For simplicity we concentrate on
  $\mathrm{Ti-Al}$ bilayers and $\mathrm{Al-Ti-Al}$ trilayers
  although the method would apply for any material combinations that can be described by BCS superconductivity. We compare the results of these proximity structures with that of Al, since the properties of Al are well characterised, both as a BCS superconductor, and as a mature KID system (e.g. NIKA). \cite{George_2007,Monfardini_2011}

 In Sec.\ref{sec:theory}
 we outline the theoretical basis of the model.
This allows us to calculate the  spatial variation of the  superconducting order parameter $\Delta(x)$, the quasiparticle and pair densities of states, and the superconducting energy gap $\Delta_{\mathrm{g}}$. We show how to calculate the complex conductivity $\sigma(x)$ and the surface impedance $Z_{\mathrm{s}}$ taking account of spatially varying film properties. We discuss how to take account of the ordering of a multilayer with respect to the incident field. From $Z_{\mathrm{s}}$ we calculate the surface impedance quality factor $Q_{\mathrm{s}}$.
We calculate the position and energy dependent quasiparticle recombination time $\tau_{\mathrm{r}}(x,E)$ ($E$ is the energy) and introduce expressions for energy- and multilayer-averaging of $\tau_{\mathrm{r}}$.
Throughout these calculations we take full account of the spatial variation of the material parameters
$T_{\textrm{c}}$, $N_0$, normal-state conductivity $\sigma_{\mathrm{N}}$, and characteristic quasiparticle recombination time
$\tau_0$ for dissimilar materials. 
Whilst some of these equations have been described previously, here we present a consistent formulation using a common
 framework. This is essential and means that the model described reduces to more usual expressions in the homogeneous BCS limit.

Section~\ref{sec:Method} gives details of the  numerical method and material parameters and we discuss the temperature scaling used to ensure that calculations of the properties of multilayers with different $\nu_{\mathrm{g}}$  are done in comparable temperature regimes. In Sec.~\ref{sec:results}
 we show some of the most important outcomes of the modeling.
 The  quasiparticle and pair densities of states calculated here deviate significantly from the homogeneous (BCS) case---this deviation is a characteristic result for any multilayer.
Section \ref{sec:conclusions} discusses and summarizes the work.
Taking proper account of spatial variation not only leads to  predictions for multilayer performance for $Q$ and $\tau_{\mathrm{r}}$ that are surprising at first sight, but also predicts significant \emph{benefits} from using multilayers as detectors operating below
$90\,\,\textrm{GHz}$.

\section{Theory}
\label{sec:theory}
\subsection{Usadel equation and self-consistency}
\noindent
Our analysis of multilayers is based on Usadel's model for the generalized proximity effect in the dirty-limit.
Layer boundaries within multilayer structures function as scattering centers and ensure that the behavior of the structure is governed by dirty-limit equations. \cite{Brammertz2001} The one-dimensional Usadel equation states that\cite{Usadel1970,Brammertz2001,Golubov2004,Vasenko2008}
\begin{equation}
\label{eq:usadel_raw}
\hbar D_S \frac{\partial }{\partial x} \left( \hat{G} \frac{\partial }{\partial x} \hat{G} \right) = [\hbar\omega_{n} \sigma_z + \hat{\Delta}(x),\hat{G}].
\end{equation}
%
The matrices are defined such that
\begin{equation}
\hat{G}(x,\hbar\omega_{n})= \begin{pmatrix}
G & F \\
F^* & -G
\end{pmatrix}\,\,{\textrm{and}}\,\,\hat{\Delta}(x)= \begin{pmatrix}
0 & \Delta(x) \\
\Delta^*(x) & 0
\end{pmatrix}, \notag
\end{equation}
where $\hbar=h/2\pi$, subscript S denotes the layer of the material, $\omega_{n}=2\pi k_{\mathrm {B}} T\left(n+\frac{1}{2}\right)/\hbar$ are the Matsubara frequencies, $n=0,1,2\dots$, $\sigma_{z}$ is the Pauli matrix, $\Delta(x)$ is the superconducting order parameter, and $[a,b]=ab-ba$ is the commutator. $D_{S}=2\pi\xi_{S}^2T_{\mathrm{c}}k_{\mathrm{B}}/\hbar$ is the diffusion coefficient, where $\xi_S$ is the coherence length.  
$G$ and $F$ are the normal and anomalous Green's functions and the normalization conditions are $\hat{G}^2=\hat{1}$, and $G^2+FF^*=1$.
The analytic continuation of the Green's functions into the continuous domain of quasiparticle energies $E$ is achieved by $\hbar\omega_{n}\rightarrow-iE$ in Eq.~(\ref{eq:usadel_raw}).

In the $\theta$-parameterization, which automatically satisfies the normalization conditions, we have
$$\hat{G}(x,\hbar\omega_{n})= \begin{pmatrix}
\cos\theta & \sin\theta \\
\sin\theta & -\cos\theta
\end{pmatrix},$$
where $\theta = \theta(x,\hbar\omega_{n})$. This yields the $\theta$-parameterized Usadel equation
\begin{equation}
\frac{\hbar D_S}{2} \pdv[2]{\theta}{x}=\hbar\omega_{n} \sin \theta - \Delta(x) \cos \theta.
\label{eq:usadel}
\end{equation}
%
Eq.~\ref{eq:usadel} needs to be solved along with the 
 superconducting self-consistency equation
\begin{equation}
\Delta(x) \ln \left(\frac{T_{\mathrm{c}}}{T} \right) - 2 \pi k_{\mathrm {B}} T \sum_{\omega_{n} > 0} \left( \frac{\Delta(x)}{\hbar\omega_{n}} - \sin\theta \right) = 0,
\label{eq:selfCon}
\end{equation}
subject to appropriate boundary conditions (BCs) introduced in Ref. \citenumns{Kuprianov1988}.  At the open boundaries of the $\mathrm{S}$ or $\mathrm{S'}$ layers of bilayer $\mathrm{S'-S}$ or trilayer $\mathrm{S-S'-S}$ structures, the BCs are
\begin{align}
 \pdv{\theta_{S,S'}}{x}=0.
 \label{eq:BC-open}
\end{align}
At the $\mathrm{S-S'}$ interfaces,
\begin{align}
 \frac{1}{\rho_{S}}\pdv{\theta_{S}}{x}=\frac{1}{\rho_{S'}}\pdv{\theta_{S'}}{x},
 \label{eq:BC-inter1}
\end{align}
\begin{align}
\gamma_{\mathrm{B},S}\xi_{S}\pdv{\theta_{S}}{x}=\sin(\theta_{S'}-\theta_{S}).
\label{eq:BC-inter2}
\end{align}
At the $\mathrm{S'-S}$ interfaces,
\begin{align}
\gamma_{\mathrm{B},S'}\xi_{S'}\pdv{\theta_{S'}}{x}=\sin(\theta_{S}-\theta_{S'}).
\label{eq:BCUsadel_2}
\end{align}
Here $\gamma_{\mathrm{B},S}=R_{\mathrm{B}}/(\rho_{S}\xi_{S})$ is a measure of boundary resistivity, $\rho_{S,S'}$ are the normal-state resistivities of $\mathrm{S,S'}$ layer, and $R_{\mathrm{B}}$ is the product of the boundary resistance between the $\mathrm{S-S'}$ layers and its area.

Finally, the quasiparticle density of states is given by $N(x,E)=N_{0}(x)Q(x,E)$, where $Q(x,E)=\operatorname{Re}[\cos\theta(x,E)]$, and $N_{0}(x)$ is the position-dependent single spin density of electron states. We define the pair-breaking threshold $2\Delta_{\mathrm{g}}$ for the proximity structure as twice the energy for which the DoS becomes appreciably non-zero ($N/N_{0}>\delta$), where $\delta$ is the numerical precision.

\subsection{Nam's formulation for conductivities}
\noindent Nam's equations\cite{Nam1967} are a generalization of the Mattis-Bardeen\cite{MattisBardeen} theory into strong-coupling and impure superconductors. The real and imaginary parts of the complex conductivity $\sigma = \sigma_1-j\sigma_2$ can be expressed as
\begin{align}
\label{eq:con1}\frac{\sigma_1(\nu)}{\sigma_{\mathrm{N}}} = & \frac{1}{h\nu} \int_{\Delta_{\mathrm {g}}-h\nu}^{-\Delta_{\mathrm {g}}} dE \, g_1(E,h\nu+E) \\
& \phantom{{}=1}\phantom{{}=1} \left[ 1-2f(E+h\nu,T) \right] \notag \\
+ &\frac{2}{h\nu} \int_{\Delta_{\mathrm {g}}}^{\infty} dE \, g_1(E,h\nu+E) \notag \\
 & \phantom{{}=1}\phantom{{}=1} \left[ f(E,T)-f(E+h\nu,T) \right] \notag, \\
\label{eq:con2} \frac{\sigma_2(\nu)}{\sigma_{\mathrm{N}}} = & \frac{1}{h\nu} \int_{\Delta_{\mathrm {g}}-h\nu,-\Delta_{\mathrm {g}}}^{\Delta_{\mathrm {g}}} dE \, g_2(E,h\nu+E) \\
& \phantom{{}=1}\phantom{{}=1} \left[ 1-2f(E+h\nu,T) \right] \notag \\
+ &\frac{1}{h\nu} \int_{\Delta_{\mathrm {g}}}^{\infty} dE \,  \notag \\
&\phantom{{}=1} \left\{ g_2(E,h\nu+E) \left[ 1-2f(E+h\nu,T) \right] \right. \notag \\
& \phantom{{}=1}\phantom{{}=1} + \left. g_2(h\nu+E,E)\left[ 1-2f(E,T) \right] \right\} \notag,
\end{align}
where the lower limit of the first integral in Eq.~(\ref{eq:con2}) refers to the larger of the two
energies $\Delta_{\mathrm {g}}-h\nu$ and $-\Delta_{\mathrm {g}}$,
 and  $f(E,T)$ is the Fermi distribution function.
The coherence factors $g_{1,2}$ are given by
\begin{align}
& g_1(E,h\nu+E)=Q(E)Q(h\nu+E)+P(E)P(h\nu+E), \notag \\
& g_2(E,h\nu+E)=-\tilde{Q}(E)Q(h\nu+E)-\tilde{P}(E)P(h\nu+E), \notag
\end{align}
where $Q(E)$, $\tilde{Q}(E)$, $P(E)$, $\tilde{P}(E)$, are the generalized quasiparticle and pair densities of states
\begin{align}
& Q(E)+j \tilde{Q}(E)=\operatorname{Re}[\cos\theta(E)]+j \operatorname{Im}[\cos\theta(E)] \notag \\
& P(E)+j \tilde{P}(E)=\operatorname{Re}[-j\sin\theta(E)]+j \operatorname{Im}[-j\sin\theta(E)]. \notag
\end{align}
\subsection{Surface Impedance, Transfer Matrices and Quality Factor}
\begin{figure}[h]
\includegraphics[width=8.6cm]{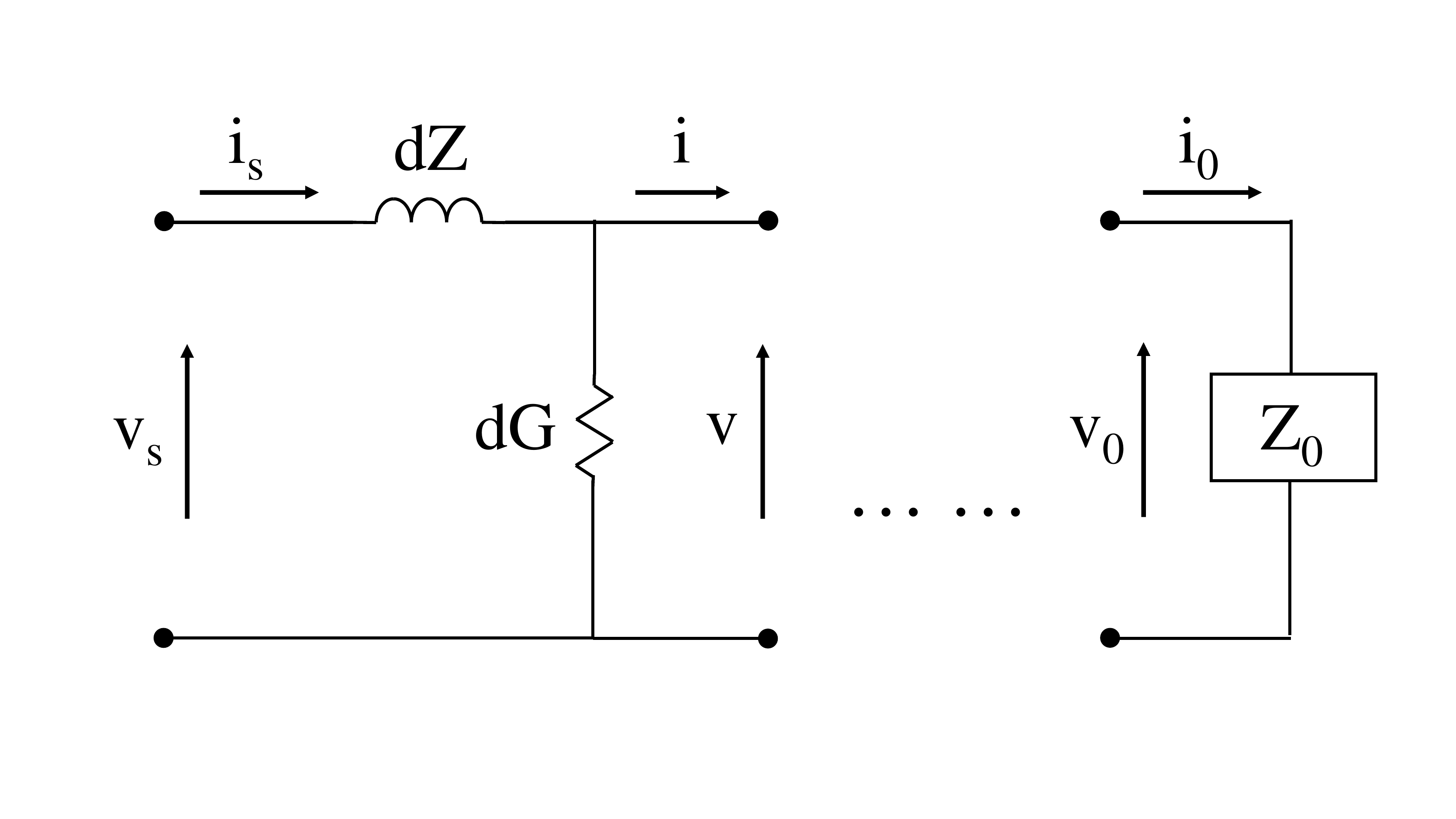}
\caption{\label{fig:transferDiag} Transmission line representation of a multilayer as a combination of layers of thickness $dx$, terminated at vacuum of impedance $Z_{0}$.}
\end{figure}
In order to calculate $Zs$, we use a transmission line model where the multilayer is sub-divided into a combination of thin layers\cite{Kerr1996} each represented by an equivalent circuit of series impedance $dZ=j 2\pi\nu \mu_0 dx$ and shunt admittance $dG=\sigma(\nu,x)dx$, terminating at vacuum, as shown in Fig. \ref{fig:transferDiag}. We represent  the
 thin layers by transmission (ABCD) matrices.\cite{KChang1994} Cascading the resulting matrices of all layers
we have
\begin{equation}
\begin{bmatrix} v_{\mathrm{s}}\\ i_{\mathrm{s}} \end{bmatrix}=\prod_{all \ layers} \begin{bmatrix} 1 & j2\pi\nu\mu_0 dx\\ \sigma(\nu,x)dx&1 \end{bmatrix} \begin{bmatrix} v_0\\ i_0 \end{bmatrix},
\end{equation}
where $v_{0}/i_{0}=Z_{0}$ is the impedance of free space.
The surface impedance is then $Z_{\mathrm{s}}=v_{\mathrm{s}}/i_{\mathrm{s}}$.
The layer facing the incoming field corresponds to the left-most matrix in the cascaded multiplication chain.
In this way calculation of $Z_{\mathrm{s}}$ for multilayers straightforwardly takes account of the physical ordering
 of  high and low conductivity films with respect to the incident field.
The model described  is a full calculation of $Z_{\mathrm{s}}$ and reduces to the expressions in Ref. \citenumns{Kerr1996}
 for homogeneous superconducting films in all limits.

We  quantify $Q$ for a multilayer resonator used as a KID with the surface impedance quality factor\cite{Jonas_review}  $Q_{\mathrm{s}}(\nu)=\operatorname{Im}(Z_{\mathrm{s}})/\operatorname{Re}(Z_{\mathrm{s}})$.
$Q_{\mathrm{s}}$ is representative of the achievable $Q$, capturing the underlying physics of the spatially-dependent energy storage and loss in the multilayer, and is related to $Q$ via Eq.~(20-21) of Ref.~\citenumns{Jonas_review}.
 A complete model to calculate $Q$ would require a full electromagnetic model of the KID geometry (be-it microstrip, coplanar waveguide, lumped-element KID etc.) including electromagnetic coupling and field-fringing effects, but this is beyond the scope of this work.
Our calculations of
 $Z_{\mathrm{s}}$ for superconducting multilayers will enable full electromagnetic  modeling to be done \emph{for the first time}.
\subsection{Multilayer-averaged Quasiparticle Recombination Times}
\noindent We use the low-energy expression for quasiparticle recombination time given by Ref. \citenumns{Golubov1994} and the material-characteristic electron-phonon coupling time constant $\tau_0$ as described by
 Kaplan {\textit{et al.}}.\cite{kaplan1976}
 We ignore quasiparticle scattering, which becomes very slow at low temperatures for low energy quasiparticles.\cite{kaplan1976}
 The  recombination lifetime is given by
\begin{align}
\frac{\tau_0(x)}{\tau_{\mathrm{r}}(x,E)}=&\frac{1}{(k_{\mathrm {B}}T_{\mathrm {c}})^3[1-f(E,T)]} \int_{E+\Delta_{\mathrm {g}}(x)}^{\infty} d\Omega \, \Omega^2 \notag \\
& \left[ n(\Omega,T)+1\right] f(\Omega-E,T) \notag \\
& \left( Q(x,\Omega-E)+\frac{\Delta(x)}{E}P(x,\Omega-E) \right),
\label{eq:Golubov_corrected}
\end{align}
where $\Omega$ is the phonon energy and $n(\Omega,T)$ is the Bose distribution. We note here the change in sign within the final bracket of Eq.~(\ref{eq:Golubov_corrected}) compared with
Eq.~(7d) of Ref.~\citenumns{Golubov1994}.
This ensures that  $\tau_{\mathrm{r}}$ here reduces to Kaplan's result (Eq.~(8), Ref.~\citenumns{kaplan1976}) with the appropriate expressions for $Q$ and $P$ for homogeneous BCS superconductors.

Assuming thermal equilibrium, we use $Q(x,E)$ and $f(E,T)$ as weights  to obtain the energy-averaged recombination time as a function of position
\begin{equation}
\label{eq:tau_r_1}
\langle\tau_{\mathrm{r}}(x)\rangle_{E}=\frac{\int_{\Delta_{\mathrm {g}}(x)}^{\infty} \tau_{\mathrm{r}}(x,E)N_0(x)Q(x,E)f(E,T)\,dE}{\int_{\Delta_{\mathrm {g}}(x)}^{\infty} N_0(x)Q(x,E)f(E,T)\,dE}.
\end{equation}
Applying weighted averaging over both energy and position, the multilayer-averaged quasiparticle recombination time is obtained,
\begin{equation}
\label{eq:tau_r_2}\langle\tau_{\mathrm{r}}\rangle_{E,x}=\frac{\int\int_{\Delta_{\mathrm {g}}(x)}^{\infty} \tau_{\mathrm{r}}(x,E)N_0(x)Q(x,E)f(E,T)\,dE\,dx}{\int\int_{\Delta_{\mathrm {g}}(x)}^{\infty} N_0(x)Q(x,E)f(E,T)\,dE\,dx}.
\end{equation}
For thin films at low temperatures where $\langle\tau_{r}\rangle_{E,x}\gg d_{\mathrm{Al,Ti}}^2/2{D_{\mathrm{Al,Ti}}}$,
Eq.~\ref{eq:tau_r_2}
 represents the best-estimate of the overall quasiparticle recombination time, including the effect of quasiparticle trapping, provided the number of excess quasiparticles is small such that recombination with thermal quasiparticles dominates. A full solution for strong non-equilibrium conditions, such as high loading or strong pulses, requires solving the self-consistency equation with the kinetic equations describing quasiparticle and phonon coupling,\cite{Chang1978} and is beyond the scope of the current work.
\section{Methodology}
\label{sec:Method}
\begin{figure}[h]
     \centering
     \subfloat[]{\includegraphics[width=8.6cm]{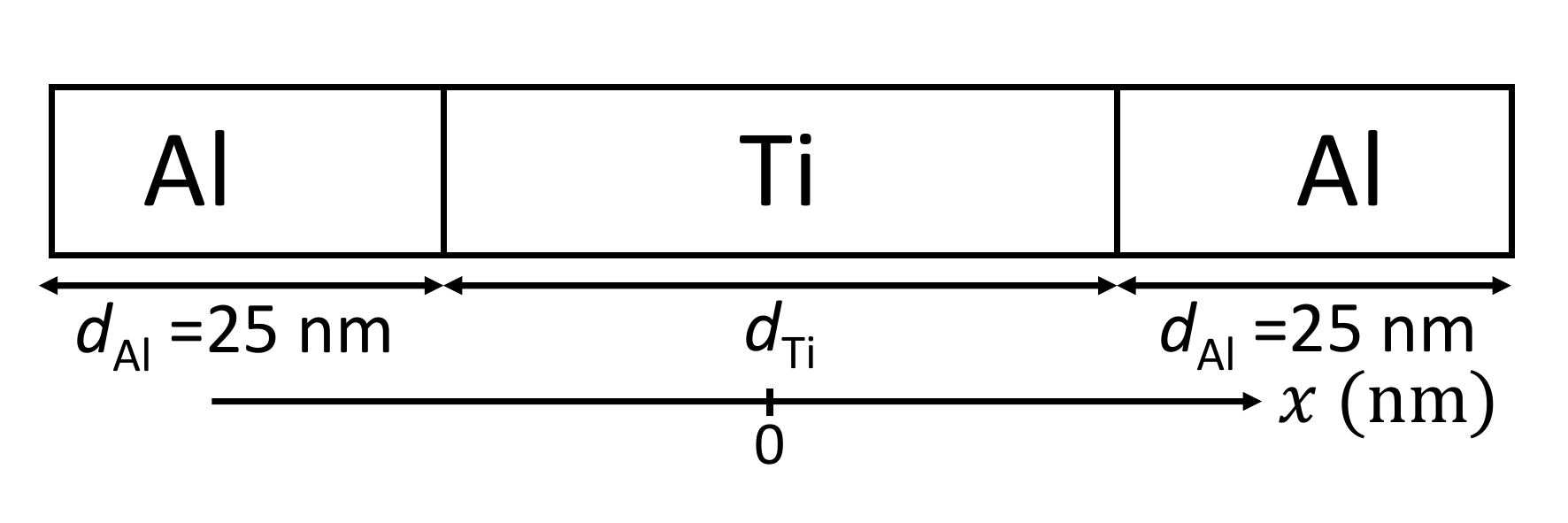}\label{fig:geometry-a}}
     \newline
     \subfloat[]{\includegraphics[width=8.6cm]{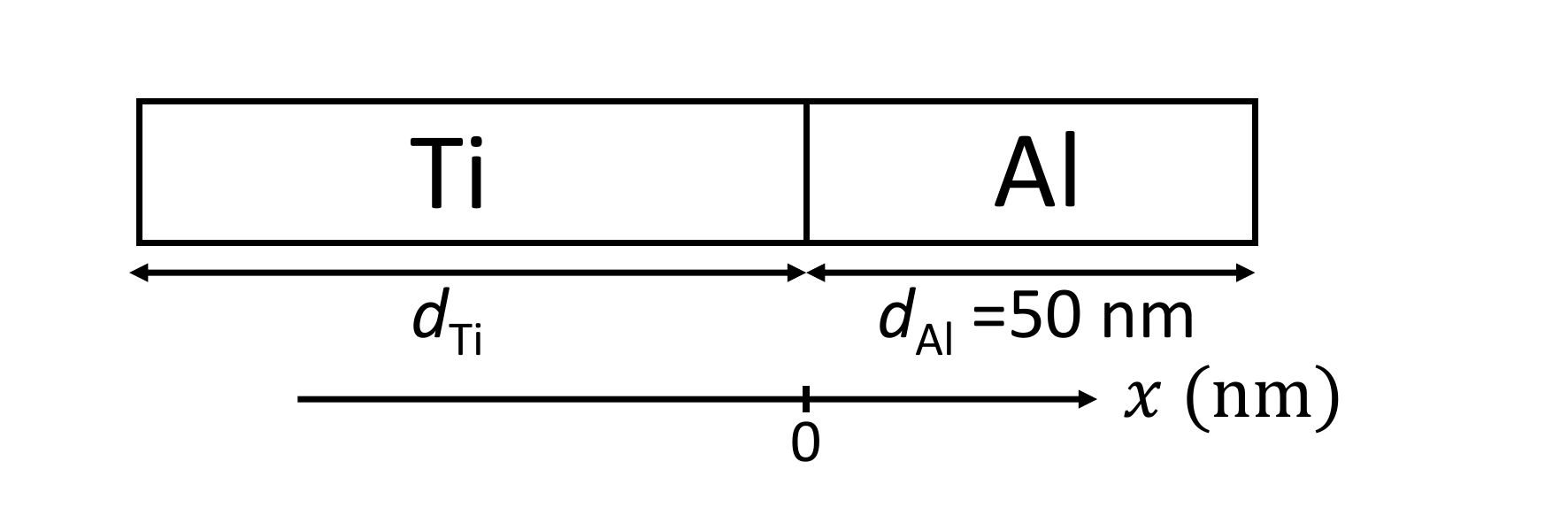}\label{fig:geometry-b}}
     \caption{(a) The geometry of trilayer $\mathrm{Al-Ti-Al}$ devices.  Al layers have $d_{\mathrm{Al}}=25\,\textrm{nm}$. The  central Ti thickness is $d_{\mathrm{Ti}}$. Position is denoted by coordinate $x$ which is zero at the centre of the Ti layer. (b) The geometry of bilayer devices. The  Al has $d_{\mathrm{Al}}=50\,\textrm{nm}$. $x=0$ coincides with the $\mathrm{Ti-Al}$ interface.}
     \label{fig:geometry}
\end{figure}
\begin{table}[ht]
\begin{threeparttable}
\caption{\label{tab:table1}Table of physical parameters of material properties.}
\begin{tabular}{b{0.50\linewidth} b{0.26\linewidth} b{0.18\linewidth}}
\toprule
 & \textrm{Aluminium} & \textrm{Titanium} \\
\colrule
$T_{\mathrm {c}}$ (K) & 1.2\tnote{1} & 0.4\tnote{1} \\
$\Delta_{\mathrm {g}}$ ($\mu eV$) & 182 & 61 \\
$\nu_{\mathrm {g}}$ (GHz) & 88 & 29 \\
$\xi$ (nm) & 170\tnote{2} & 110\tnote{2} \\
$N_0$ ($10^{23}$/$\textrm{eV}\,\textrm{cm}^3$) & 0.174\tnote{1} & 0.41\tnote{1} \\
$V$ ($10^{-23}$ $\textrm{eV}\,\textrm{cm}^3$) & 0.960\tnote{1} & 0.344\tnote{1} \\
$\Theta_D$ (K) & 423\tnote{1} & 426\tnote{1} \\
$\sigma_{\mathrm{N}}$ (/$\mu\Omega\,$m) & 180\tnote{3} & 3\tnote{4} \\
$\tau_0$ (ns) & 395\tnote{5} & 7960\tnote{5} \\
$RRR$ & 5.5\tnote{3} &  3.5\tnote{6} \\
\toprule
\end{tabular}
\begin{tablenotes}[flushleft]
\RaggedRight
\footnotesize
\item[1] Reference \citenumns{Gladstone}.

\item[2] Calculated using dirty limit expression for $\xi$ from Ref. \citenumns{Brammertz2001}.

\item[3] Measurements by the Cambridge Quantum Sensors Group.

\item[4] Reference \citenumns{Fujii2012}.

\item[5] Reference \citenumns{Parlato2005}.

\item[6] Reference \citenumns{Fukuda2007}
\end{tablenotes}
\end{threeparttable}
\end{table}
\noindent The Usadel equation (\ref{eq:usadel}) and the self-consistency equation (\ref{eq:selfCon}) are solved iteratively, with appropriate BCs, for $\mathrm{Ti-Al}$ bilayers and $\mathrm{Al-Ti-Al}$ trilayers with the geometry shown in Fig.~\ref{fig:geometry}, at $T=0.1\,\mathrm{K}$ and $\gamma_{\mathrm{B,Al}}=0.01$ or $\gamma_{\mathrm{B,Al}}=100$ to represent both high and low transmission interfaces. Our solver uses MATLAB's multipoint boundary value problem numerical solver. The solver computes the residuals using geometry-specific boundary function handles, to ensure that the solution satisfies Eqs.~(\ref{eq:BC-open}) to (\ref{eq:BCUsadel_2}). The detailed steps of our implementation are as follows:
\begin{enumerate}
    \item[(i)] A uniform trial $\Delta(x)=(\Delta_{\mathrm{Al}}+\Delta_{\mathrm{Ti}})/2$ is used to solve for $\theta(x,\hbar\omega_{n})$, using Eq.~(\ref{eq:usadel}) with maximum $n=125$. $\theta(x,\hbar\omega_{n})$ is then used to solve for $\Delta(x)$ using Eq.~(\ref{eq:selfCon}). The solution then updates $\Delta(x)$.
    \item[(ii)] An iterative loop of step (i) continues until convergence is achieved in $\Delta(x)$, such that the relative change in $\Delta(x)$ between the final two iterations is less than 0.001\%.
    \item[(iii)] The Usadel equation is then solved \textit{once} using $\Delta(x)$ from (ii) in the domain of continuous quasiparticle energy $E$ to obtain $\theta(x,E)$. This is done by applying $\hbar\omega_{n}\rightarrow-iE$ to Eq.~(\ref{eq:usadel}).
    \item[(iv)] Other properties, i.e. $\sigma_{1,2},Z_{\mathrm{s}},Q_{\mathrm{s}},$ and $\tau_{\mathrm{r}}$, are calculated from $\theta(x,E)$ using equations described in Sec.~\ref{sec:theory}.
\end{enumerate}

A list of basic physical parameters used in this paper can be found in Table \ref{tab:table1}. Data for superconducting coherence lengths are calculated using the dirty limit expression, $\xi=\sqrt{\xi_{0}l/3}$, from Ref. \citenumns{Brammertz2001}, where $\xi_{0}$ is the BCS coherence length, and $l$ is the electron mean free path. Calculation of $\xi$ is supplemented by Al residual resistivity ratio ($RRR_{\mathrm{Al}}$) from the Cambridge Quantum Sensors Group, Ti residual resistivity ratio ($RRR_{\mathrm{Ti}}$) from Ref.~\citenumns{Fukuda2007}, and Ti mean free path at 300K ($l_{\mathrm{Ti,300 K}}$) from Ref.~\citenumns{Reale1973}.
\subsection{Scaling the temperature}
\label{subsec:Scaling}
\noindent We scale the temperature of calculation for both $Q_{\mathrm{s}}$ and $\tau_{\mathrm{r}}$, in order to
remove the effect of reducing $\nu_\mathrm{g}$.
Reducing the temperature of the calculation  removes  the effect of a reduced $\Delta_{\mathrm{g}}$ to first-order.
The principal temperature dependence of both $Q_{\mathrm{s}}$ and $\tau_{\mathrm{r}}$  arises from the Fermi functions in
Eqs.~\ref{eq:con1}, \ref{eq:con2} and
\ref{eq:tau_r_1}
leading to a close-to $\exp(-\Delta_{\mathrm{g}}/k_{\mathrm{B}} T)$ dependence at a \textit{fixed}  $T$ in both cases. Scaling $T$ removes this effect, giving comparable operating regimes but it also means that in practice lower experimental temperatures are required.
 For Al, we choose $T=170\,\,\textrm{mK}$ i.e. close to  the onset of the experimentally-observed low-temperature limit of $\tau_{\mathrm{r}}$
  seen in all low-$T_\mathrm{c}$ superconductors .\cite{Jonas_review,Barends_tau_Q_2007,de_Visser_2014}
The limiting temperature scales as $T/T_\mathrm{c}$ for many low temperature superconductors, although we are unaware of similar measurements in multilayers.
 For all multilayers,
 the temperature is scaled such that $T = 170\,\,\textrm{mK}\times (\Delta_{\mathrm {g}}/\Delta_{\mathrm {Al}})$.
Even-so, for the thickest trilayer with threshold as low as    $\nu_{\mathrm{g}}= 50 \,\,\textrm{GHz}$ an operating temperature $T=95\,\,\textrm{mK}$ would be sufficient, well-above the experimentally accessible range using an ADR.

\section{Results}
\label{sec:results}
\subsection{Density of states gap and pair-breaking threshold}
\label{subsec:DoS}
\begin{figure}[h]
     \includegraphics[width=8.6cm]{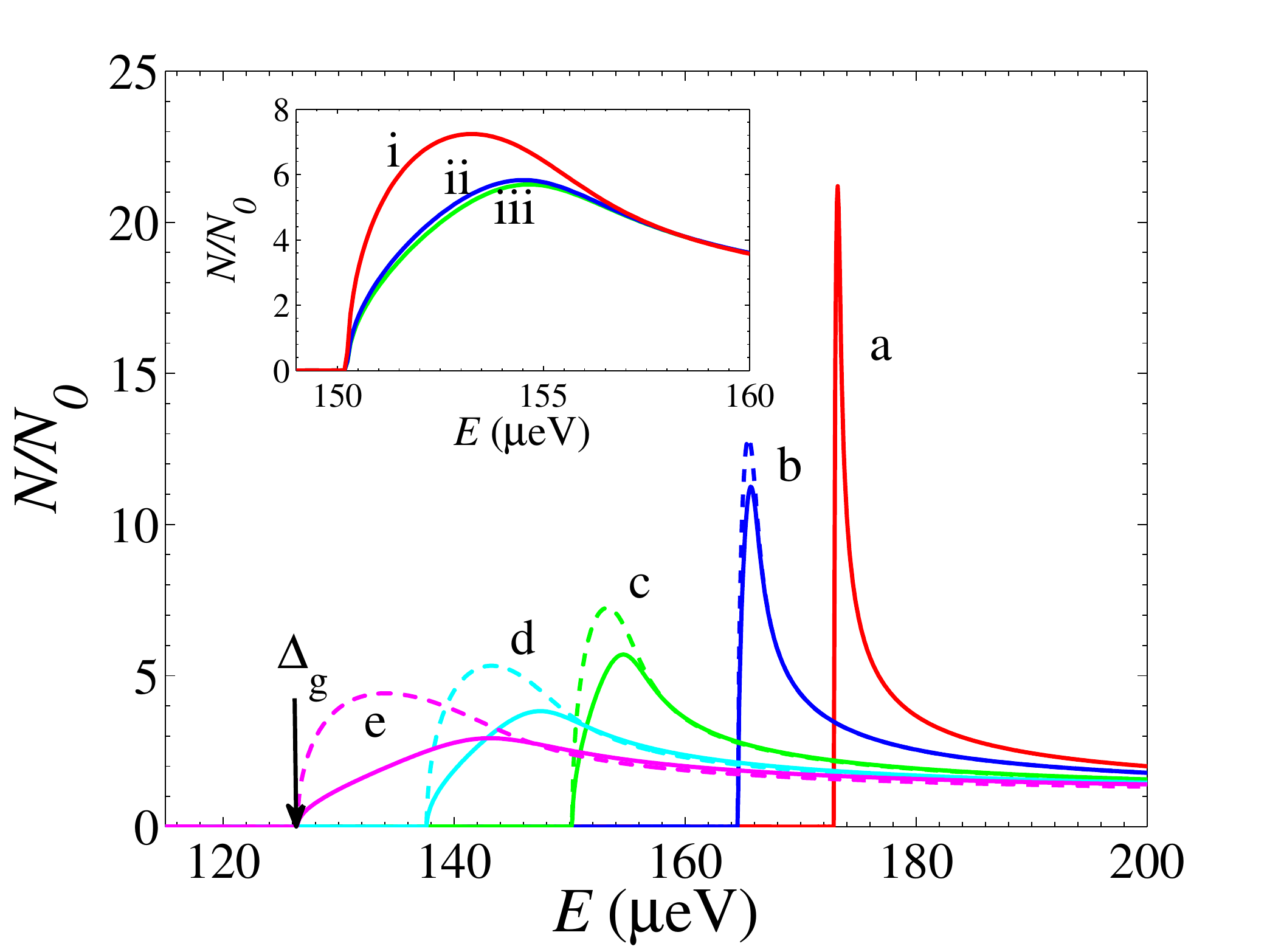}
     \caption{DoS for 5 $\mathrm{Al-Ti-Al}$ \emph{trilayers} at the open boundary of the Al (solid lines) and at the centre of the Ti (dashed lines). (a) red line,~$d_{\mathrm{Ti}}=25\,\textrm{nm}$, (b) blue line,~$d_{\mathrm{Ti}}=50\,\textrm{nm}$, (c) green line,~$d_{\mathrm{Ti}}=100\,\textrm{nm}$, (d) cyan line,~$d_{\mathrm{Ti}}=150\,\textrm{nm}$, and (e) purple line,~$d_{\mathrm{Ti}}=200\,\textrm{nm}$. The total Al thickness is $50\,\,{\textrm{nm}}$, $T=0.1\,\,{\textrm K}$ and $\gamma_{\mathrm{B,Al}}=0.01$. The inset shows the DoS  at different positions of a $\mathrm{25 nm-100 nm-25 nm}$ multilayer. (i) red line, centre of Ti layer, (ii) blue line, Ti-side of internal Ti-Al interface and, (iii) green line, external Al boundary . $T=0.1\,\,{\textrm K}$ and $\gamma_{\mathrm{B,Al}}=0.01$.
     \label{fig:DoS2_a}}
\end{figure}
\begin{figure}[h]
     \includegraphics[width=8.6cm]{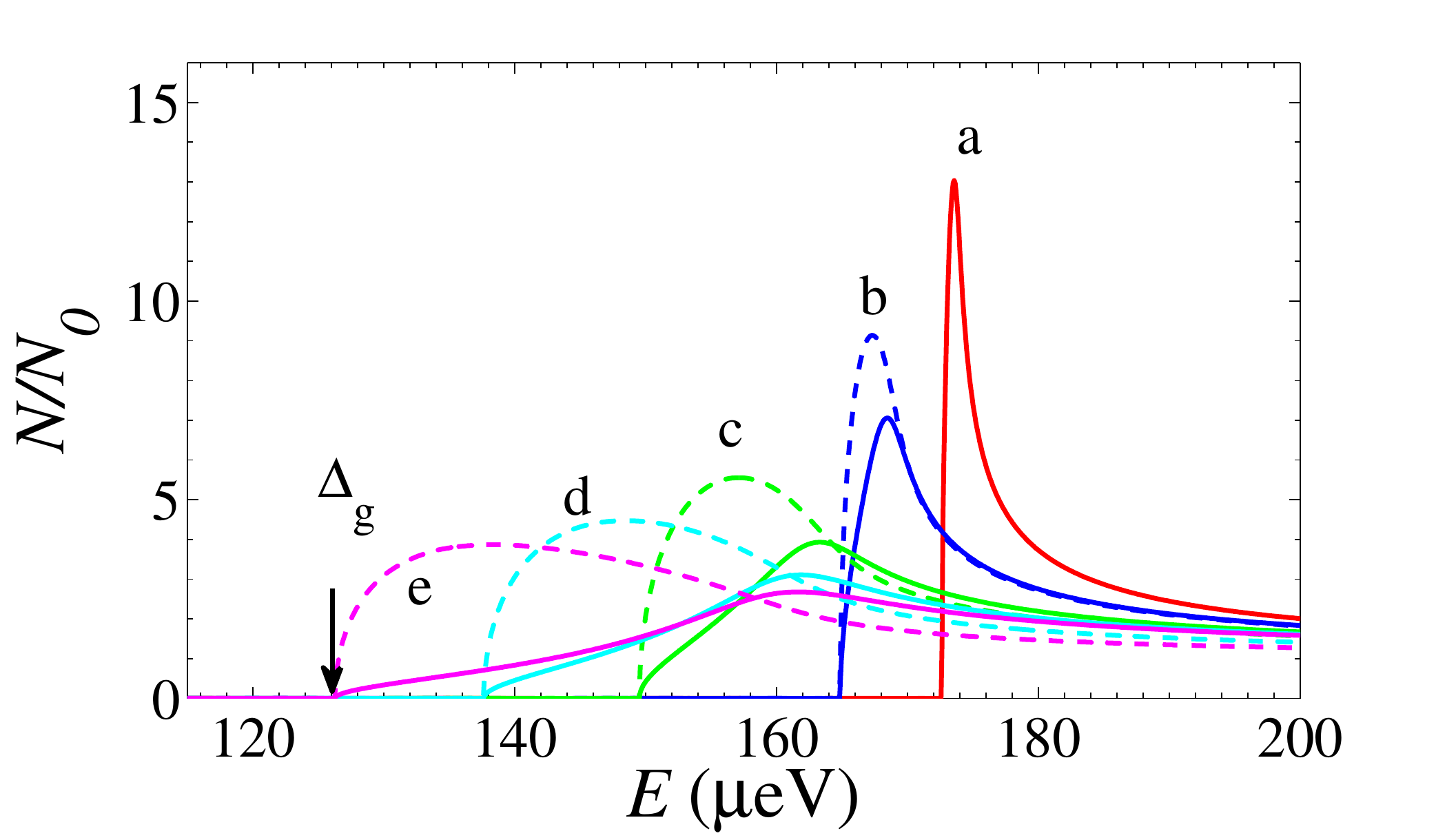}
     \caption{DoS for 5 $\mathrm{Ti-Al}$ \emph{bilayers} at the open boundary of the Al (solid lines) and at the open boundary of the Ti (dashed lines). (a) red line,~$d_{\mathrm{Ti}}=25\,\textrm{nm}$, (b) blue line,~$d_{\mathrm{Ti}}=45\,\textrm{nm}$, (c) green line,~$d_{\mathrm{Ti}}=80\,\textrm{nm}$, (d) cyan line,~$d_{\mathrm{Ti}}=105\,\textrm{nm}$, and (e) purple line,~$d_{\mathrm{Ti}}=130\,\textrm{nm}$. The total Al thickness is $50\,\,{\textrm{nm}}$, $T=0.1\,\,{\textrm K}$ and $\gamma_{\mathrm{B,Al}}=0.01$.
     \label{fig:DoS2-b}}
\end{figure}
\noindent Figure~\ref{fig:DoS2_a}  shows the normalized densities of states (DoS) for 5 $\mathrm{Al-Ti-Al}$ trilayers with Ti thicknesses, (a) to (e), 25, 50, 100, 150 and $200\,\,\mathrm{nm}$ respectively, at the open Al boundaries (solid lines) and  at the centre of the Ti layer (dashed lines). The total Al thickness is $50\,\,{\textrm{nm}}$.
  The DoS is BCS-like for the thinnest Ti layer (trace (a)) due to both $d_{\mathrm{Al}}$ and $d_{\mathrm{Ti}}$ being much smaller than their respective coherence lengths. The DoS broadens for thicker Ti films. For very thick Ti layers (not shown) the DoS at the Ti layer tends to the BCS DoS for Ti, whereas the DoS at edge of the Al broadens further and does not tend to any BCS limit. The broadening behaviour highlights the need for a rigorous model beyond BCS-like DoS approximations. In this numerical study, we use $\delta=0.01$ to determine the pair-breaking threshold $2\Delta_{\mathrm{g}}$. For clarity we have only indicated this for trace (e).
The inset  shows the DoS at different positions within the $\mathrm{25 nm-100 nm-25 nm}$ multilayer. Trace (i) shows the DoS at the centre of Ti layer, (ii) the DoS at the Ti-side of the internal Ti-Al interface, (iii) the DoS at the external Al boundary.
The DoS does not vary greatly with position in the Al layer. The high Al normal-state conductivity, compared to Ti, restricts the extent of variation of DoS in the Al layer. The DoS at different positions along the same device share a common energy gap because all films are relatively thin ($d\sim\xi$).

Figure~\ref{fig:DoS2-b} shows the DoS for 5 $\mathrm{Ti-Al}$ bilayers of Al thicknesses,
  (a) to (e), 25, 45, 80, 105 and $130\,\,\mathrm{nm}$ respectively at the open Al boundaries (solid lines) and  at the open Ti boundaries (dashed lines). Ti thicknesses for the bilayers were chosen to give the same pair-breaking thresholds as the  trilayers. A key difference between bi- and trilayers with equal $\Delta_{\mathrm{g}}$ is the reduction 
in the  DoS in the Al layer near $\Delta_{\mathrm{g}}$ for bilayers. At large values of $d_{\mathrm{Ti}}$, the reduction is more pronounced as individual Al layer thicknesses (instead of total Al layer thickness) becomes more important to the DoS. This reduction in the DoS has immediate impact on
  the thermal quasiparticle density $N_{\textrm{th}}(x)=4\int N_0 Q(x,E) f(E,T)\, dE$ in the Al and likewise changes
  $\sigma$ and $\tau_{\textrm{r}}$.
\begin{figure}[h]
\includegraphics[width=8.6cm]{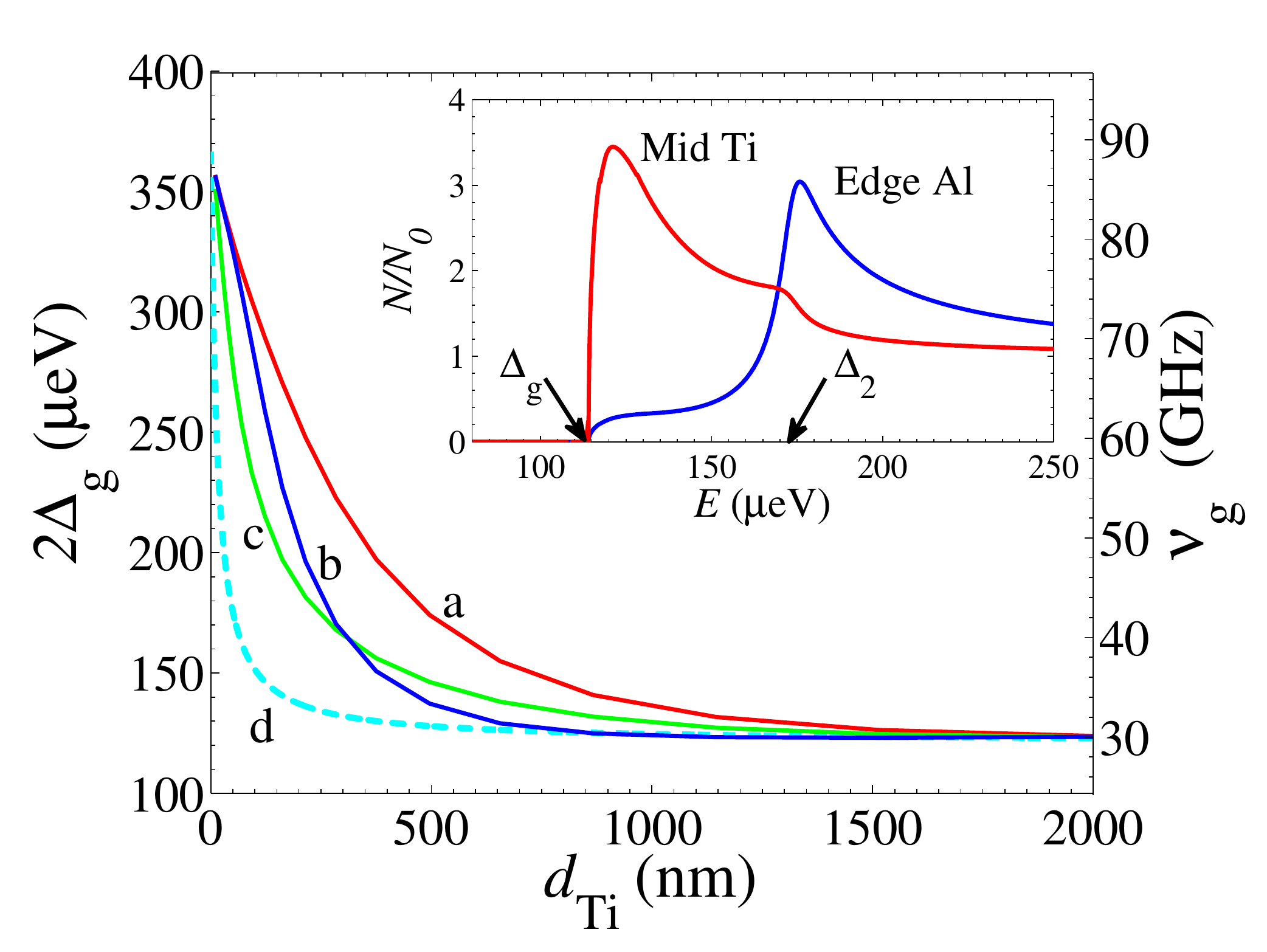}
\caption{\label{fig:Gap} Pair-breaking threshold $2\Delta_{\mathrm {g}}$ as a function of central Ti thickness with fixed total Al thickness $50\,\,\textrm{nm}$. (a) red line, trilayer with $\gamma_{\mathrm{B,Al}}=0.01$, (b) blue line, bilayer with $\gamma_{\mathrm{B,Al}}=0.01$, (c) green line, trilayer with $\gamma_{\mathrm{B,Al}}=100$, and (d) cyan line, weighted-average model. $T=0.1\,\,{\textrm K}$. The inset shows the DoS for trilayer of $\mathrm{25 nm-100 nm-25 nm}$ geometry at middle of Ti layer (red) and edge of Al layer (blue), $\gamma_{\mathrm{B,Al}}=100$.}
\end{figure}

Figure~\ref{fig:Gap} shows the pair-breaking threshold $2\Delta_{\mathrm {g}}$ (the right-hand scale shows $\nu_{\mathrm {g}}$) for: (a) red line, $\mathrm{Al-Ti-Al}$ trilayer with $\gamma_{\mathrm{B,Al}}=0.01$, (b) blue line, $\mathrm{Ti-Al}$ bilayer with $\gamma_{\mathrm{B,Al}}=0.01$, (c) green line, $\mathrm{Al-Ti-Al}$ trilayer with $\gamma_{\mathrm{B,Al}}=100$, and (d) cyan dashed line, weighted-average model.
In all cases $T=0.1\,\,{\textrm K}$  and the total Al thickness is $50\,\,\textrm{nm}$.
Comparing (a) with (b) we see that the thresholds for bi- and trilayers are not the same and the bilayer gap is \textit{lower} with the same Ti thickness and interface transparency.
This is a direct result of the BCs Eqs.~(\ref{eq:BC-open}) to (\ref{eq:BCUsadel_2}) and self-consistency Eq.~(\ref{eq:selfCon}).
Al as the outer layers of a trilayer acts to increases $\Delta(x)$
in the  Ti  from both sides, whereas the external BCs for Ti in a bilayer cannot.
Comparing (a) with (c) where $\gamma_{\mathrm{B,Al}}=0.01,\,\,100$ respectively representing a clean, high-transmission or a dirty, low-transmission interface, we see that the threshold is reduced as the interface becomes less transmissive.
In other simulations we find that $\Delta_{\mathrm {g}}$ is a weak function of $\gamma_{\mathrm{B,Al}}$ provided $\gamma_{\mathrm{B,Al}} \lesssim 1 $.
Comparing all of (a)-(c) with (d) the weighted-average model, we see that the latter significantly \emph{over-estimates}  the efficiency of the proximity effect in reducing $\Delta_{\mathrm {g}}$. In this analysis, the fixed total Al thickness at $50\,\,\textrm{nm}$ results in deviation from the weighted-average model, even when coupled with very thin Ti layer. The weighted-average model gives the same prediction for \emph{both} bi- and trilayers  because the total Al thickness is fixed: the detail imposed by the geometry is lost.
It can be seen from all of (a) to (c) that the proximity effect is a long range \emph{diffusive} process that extends over  much longer distances than the component coherence lengths.

The inset shows the DoS for a trilayer of $\mathrm{25\,\,nm-100\,\,nm-25\,\,nm}$ geometry at the centre of the Ti layer (red line) and the outer boundary of the Al layer (blue line), calculated for $\gamma_{\mathrm{B,Al}}=100$.
In this case, where the film coupling is weak, the DoS across the multilayer shows more
structure with two relatively broad peaks, the first somewhat greater than $\Delta_{\mathrm {g}}$ and  the second $\Delta_{2}$  close to $\Delta_{\mathrm{Al}}$.
 The DoS at the centre of the Ti increases significantly from zero at $\Delta_{g}$, and shows a small increase at $\Delta_{2}$.
 The DoS at the edge of the Al increases slowly from zero at $\Delta_{g}$ and shows a large increase
  at $\Delta_{2}$.
  For very low transmission boundaries $\gamma_{\mathrm{B,Al}}\to\infty$, the DoS at the middle of the Ti increases exclusively at $\Delta_{g}=\Delta_{\mathrm {g,Ti}}$ whilst the DoS at the edge of the Al increases exclusively at $\Delta_{2}=\Delta_{\textrm {g,Al}}$, the  proximity coupling disappears in this case and we find a BCS-like DoS for the individual layers.

The overall effect on KID performance as a result of these spatially varying DoS with $d_{\mathrm{Ti}}$ is not obvious at first sight. In the next sections we explore the impacts on $Z_{\mathrm{s}}$, $Q_{\mathrm{s}}$ and $\tau_{\mathrm{r}}$ for bi- and trilayers.

\subsection{Conductivity, Impedance and Quality Factor}
\begin{figure}[h]
\includegraphics[width=8.6cm]{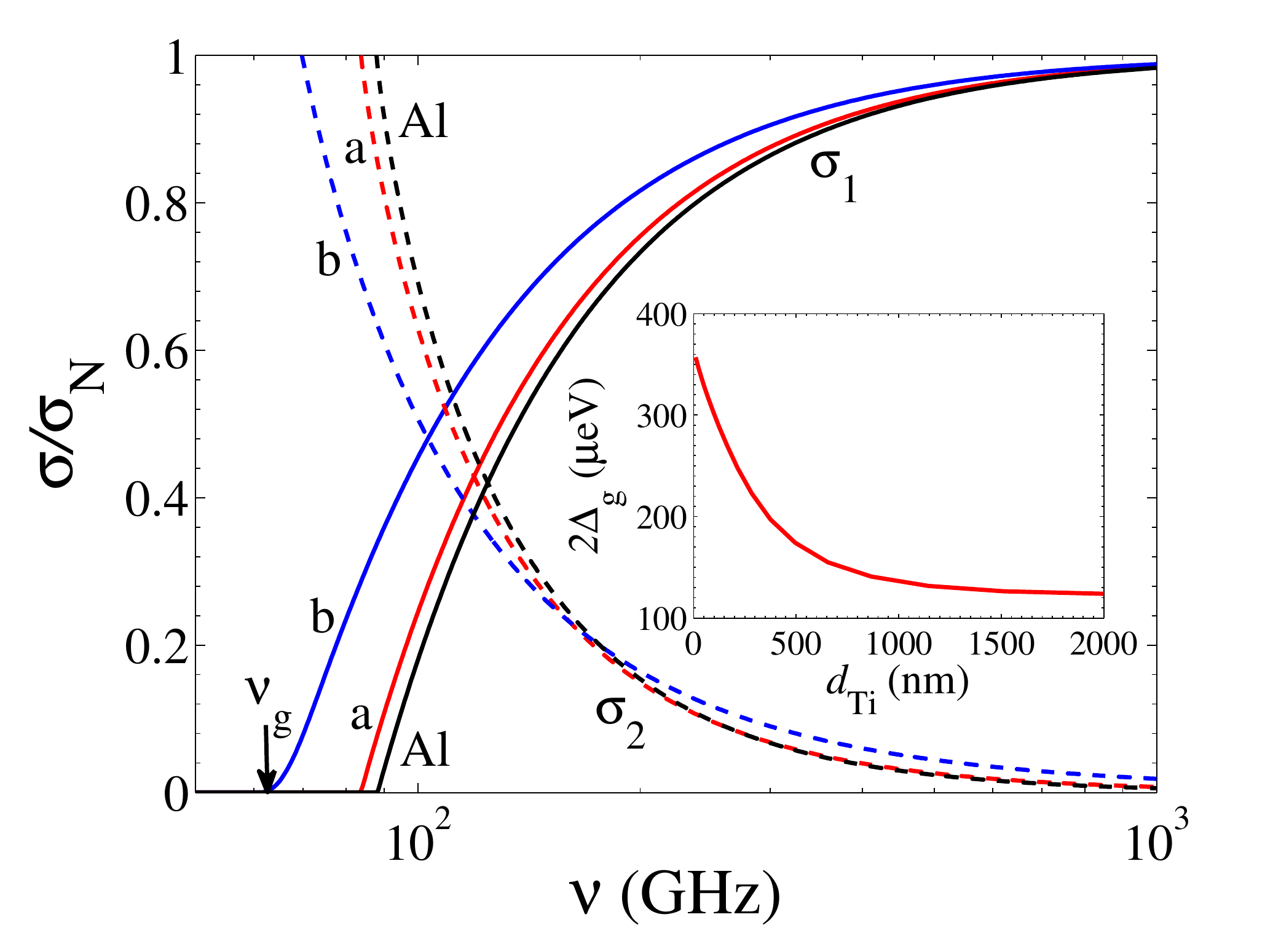}
\caption{\label{fig:Con} Normalized real (solid line) and imaginary (dashed line) conductivities for two trilayers,
(a) red line,~$d_{\mathrm{Ti}}=25\,\textrm{nm}$, (b) blue line,~$d_{\mathrm{Ti}}=200\,\textrm{nm}$. The black lines are calculated for Al. $T=0.1\,\,{\textrm K}$, and $\gamma_{\mathrm{B,Al}}=0.01$. The total Al thickness
  is $50\,\,\textrm{nm}$.  The inset shows the variation of $2\Delta_{\mathrm {g}}$ as a function of $d_{\mathrm{Ti}}$.}
\end{figure}

\noindent Figure~\ref{fig:Con} shows normalized conductivities $\sigma/\sigma_{\mathrm{N}}$ as a function of $\nu$ for two trilayers with (a)
$d_{\mathrm{Ti}}= 25\textrm{,}$ and (b) $200\,\,\textrm{nm}$.  The total Al thickness
  is $50\,\,\textrm{nm}$. Full lines show $\sigma_1$ and dashed lines $\sigma_2$. The black lines show calculations for Al for comparison.
In all cases $T=0.1\,\,{\textrm K}$ and $\gamma_{\mathrm{B,Al}}=0.01$ and $d_{\mathrm{Al}}= 50\,\,\textrm{nm}$. Also indicated is $\nu_{\mathrm{g}}$ for the $25\,\mathrm{nm}-200\,\mathrm{nm}-25\,\mathrm{nm}$ trilayer. The ratio 
$\sigma_1/\sigma_{\mathrm{N}}$ is identical to the superconducting to normal-state photon absorption ratio.\cite{Tinkham_1994}
$\sigma_1/\sigma_{\mathrm{N}}$ has a threshold at $\nu_{\mathrm {g}}=2\Delta_{\mathrm {g}}/h$ and shows a slow increase above threshold.
As a result, although Al KIDs have $\nu_{\mathrm {g}}= 88\,\,\textrm{GHz}$, they only operate efficiently above $100\,\,\textrm{GHz}$ or even $120\,\,\textrm{GHz}$.\cite{Catalano_2015}
We find that changes in the normalized $\sigma/\sigma_{\mathrm{N}}$  are small as a function of position within a particular bi- or trilayer, and thus we have not shown these results. This is a result of the highly-transmissive interfaces. However, the absolute values of $\sigma$ are very different in the Al and Ti layers due to the very different values of $\sigma_{\mathrm{N}}$.

\begin{figure}[h]
\includegraphics[width=8.6cm]{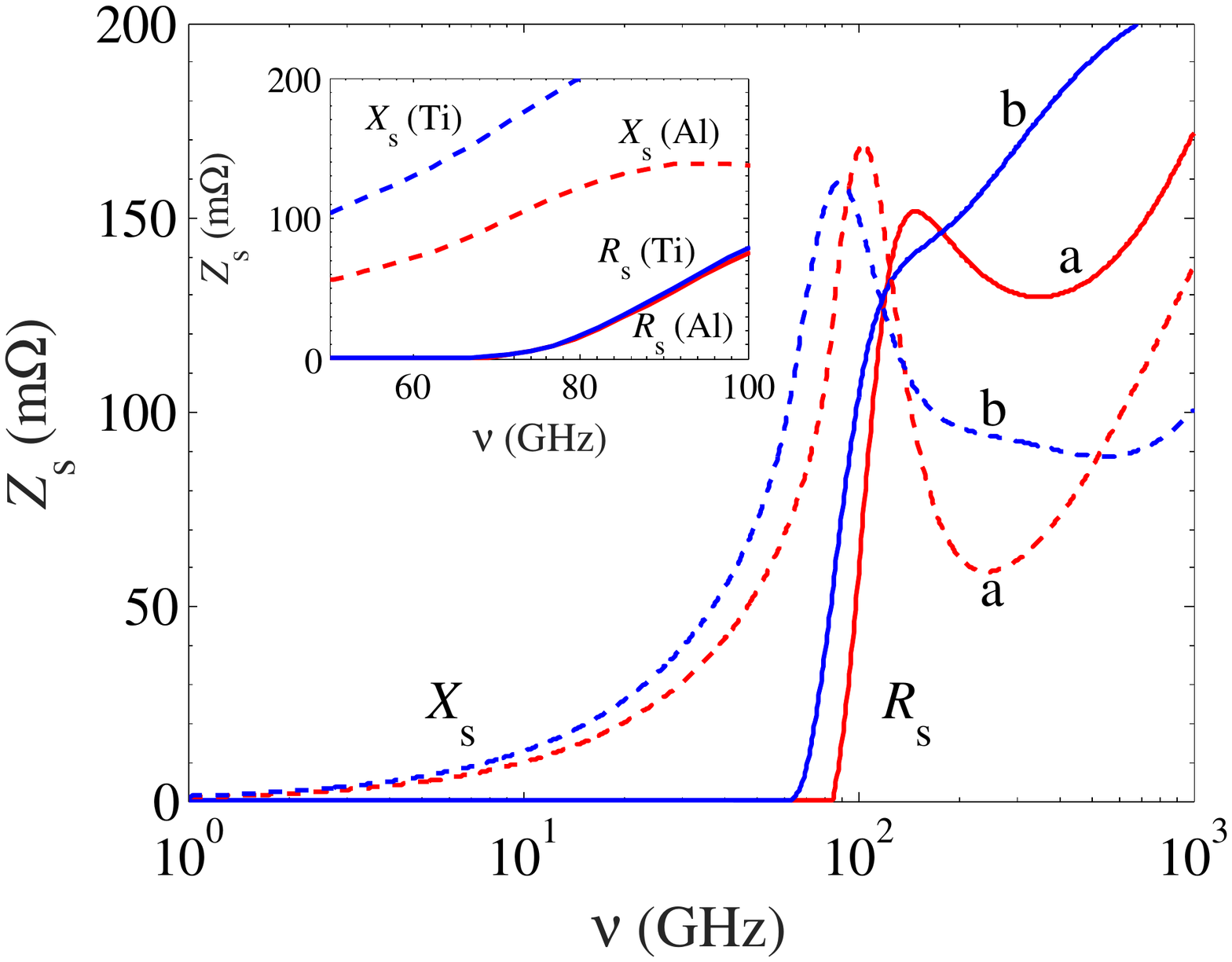}
\caption{\label{fig:Zs} Frequency dependence of $R_{\mathrm{s}}$ (solid line) and $X_{\mathrm{s}}$ (dashed line) for two trilayers:
 (a) red line,~$d_{\mathrm{Ti}}=25\,\textrm{nm}$, (b) blue line,~$d_{\mathrm{Ti}}=200\,\textrm{nm}$. $T=0.1\,\,{\textrm K}$ and $\gamma_{\mathrm{B,Al}}=0.01$ and  $d_{\mathrm{Al}}=25\,\textrm{nm}$. The inset shows the variation of $R_{\mathrm{s}}$ (solid lines) and $X_{\mathrm{s}}$ (dashed lines) for a bilayer with $d_{\mathrm{Ti}}=130\,\textrm{nm}$, at frequencies close to $\nu_{\mathrm {g}} = 61\,\mathrm{ GHz}$. The blue lines indicate incidence on the Ti surface, the red lines indicate incidence on the Al surface.} 
\end{figure}
\begin{figure}[h]
\includegraphics[width=8.6cm]{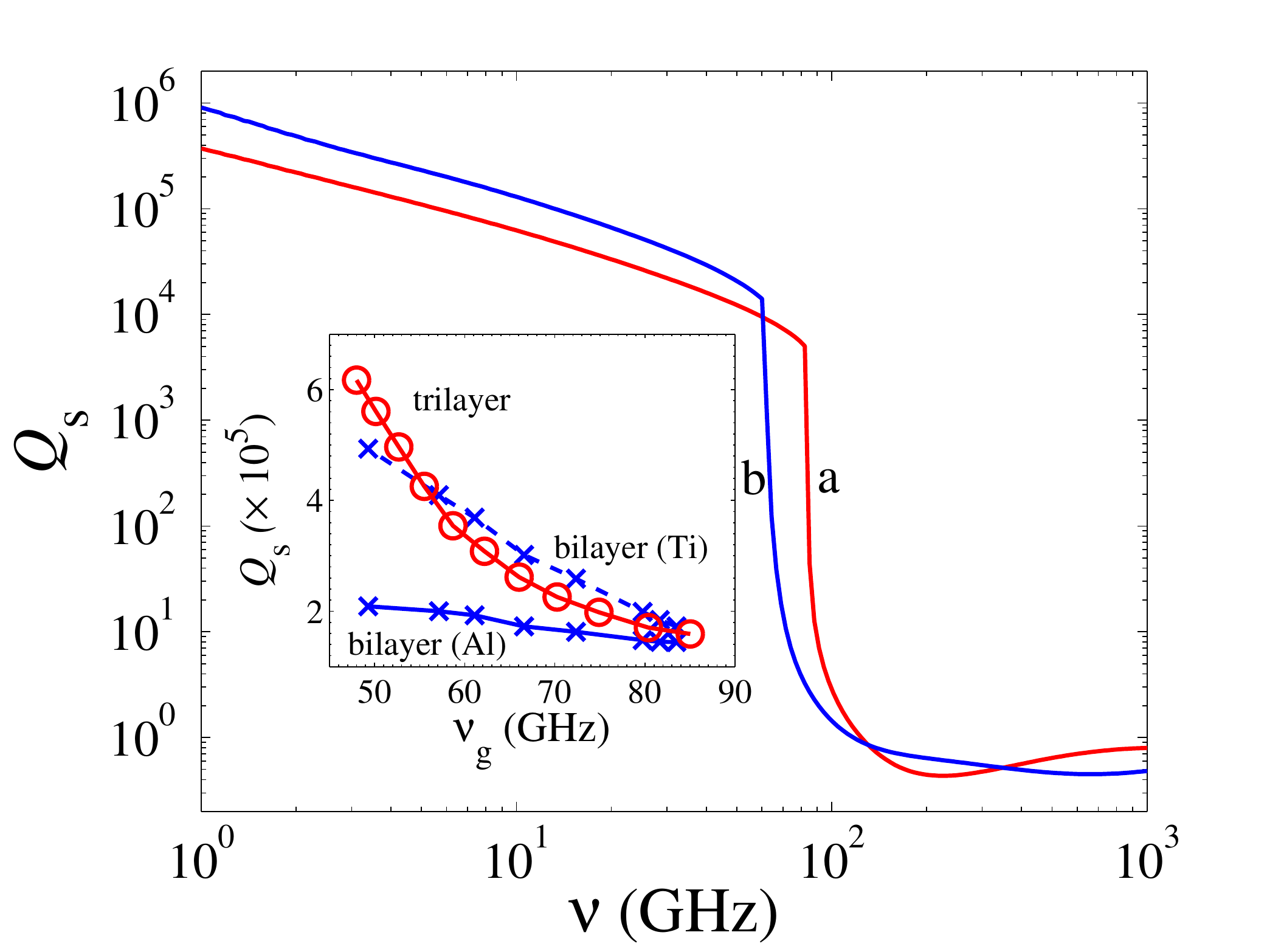}
\caption{\label{fig:Qs} Frequency dependence of $Q_{\mathrm{s}}$ for trilayers with varying $d_{\mathrm{Ti}}$. (a) red line,~$d_{\mathrm{Ti}}=25\,\textrm{nm}$, (b) blue line,~$d_{\mathrm{Ti}}=200\,\textrm{nm}$. $T=0.1\,\,{\textrm K}$, $\gamma_{\mathrm{B,Al}}=0.01$, with total  Al thickness of $50\,\,\textrm{nm}$.
The inset shows the variation of $Q_{\mathrm{s}}$ with $\nu=3\,\,\mathrm{ GHz}$
as a function of $\nu_{\mathrm {g}}$: (red solid line) trilayers, (blue solid line) bilayers with incidence on the Al surface,  (blue dashed line) bilayers with incidence on the Ti surface.
The temperature of the calculation is scaled such that
$T = 170\,\,\textrm{mK}\times (\Delta_{\mathrm {g}}/\Delta_{\mathrm {Al}})$  for all multilayers.}
\end{figure}
Figure~\ref{fig:Zs} shows the frequency dependence of the real and imaginary parts of $Z_{\mathrm{s}}=R_{\mathrm{s}}+jX_{\mathrm{s}}$ for 2 trilayers
with (a) $d_{\mathrm{Ti}}=25$, and (b) $200\,\,\textrm{nm}$.
Here $\nu$ is chosen to emphasize \emph{signal} frequencies. The structure evident in $R_{\mathrm{s}}$ and $X_{\mathrm{s}}$
 above $\nu_\textrm{g}$ arises because \emph{both} the DoS and $\sigma_{\mathrm{N}}$ are dependent on position. This structure would be absent if a simple BCS-like DoS were assumed to carry out the calculation of $R_{\mathrm{s}}$ and $X_{\mathrm{s}}$.
The surface resistance $R_{\mathrm{s}}$ decreases sharply with decreasing frequencies near $\nu_\textrm{g}$, and tends-to zero  below $\nu_\textrm{g}$ as $ \exp \left(- \nu_\textrm{g}/k_{\textrm{B}}T \right)$.
We find that contributions from the Al layers to $Z_{\mathrm{s}}$ are more important by virtue of its higher $\sigma_{\mathrm{N}}$ compared to Ti.
Interestingly, at low frequencies, the thickest trilayer (blue line) has the lowest dissipative component $R_{\mathrm{s}}$ (barely visible) and highest reactive component $X_{\mathrm{s}}$.
The inset of Fig.~\ref{fig:Zs} shows the variation of $R_{\mathrm{s}}$ and $X_{\mathrm{s}}$ for a bilayer with $d_{\mathrm{Ti}}=130\,\textrm{nm}$, at frequencies close to $\nu_{\mathrm {g}} = 61\,\mathrm{ GHz}$. $R_{\mathrm{s}}$ is almost independent of orientation, while $X_{\mathrm{s}}$ shows significant differences when a bilayer is illuminated from different sides. This highlights the need to properly account for geometry when modeling multilayer resonators.

In order to show the effect on the detection sensitivity of a KID fabricated from these bi- and trilayers, in Fig.~\ref{fig:Qs} we plot  $Q_{\mathrm{s}}$, at  scaled temperatures in the range of $100\,\,\mathrm{mK}$ for the lowest gap multilayers up to $170\,\,\mathrm{mK}$ for Al.
$Q_{\mathrm{s}}$ increases by a factor $\simeq 2$ as $d_{\mathrm{Ti}}$ increases from 25 to $200\,\textrm{nm}$. Changing the thickness of an Al KID by the same amount would change $Q_S$ by a factor $\simeq 3$.
From the point-of-view of KID operation
this begins to suggest that there is little penalty in device sensitivity even having engineered the reduced threshold.

The inset of Fig.~\ref{fig:Qs} shows $Q_{\mathrm{s}}$ at $\nu=3\,\mathrm{GHz}$ representing a typical KID readout frequency, as a function of
threshold $\nu_\mathrm{g}$. The solid blue line shows  $Q_{\mathrm{s}}$ for bilayers with the field incident on the  Al surface, the dashed blue line shows $Q_{\mathrm{s}}$ for bilayers with field incident on the  Ti surface.
In both cases as $\nu_{g}$ is reduced  $Q_{\mathrm{s}}$ increases  due to the increase in total thickness but note that the
geometry--- in this case the ordering with respect to the incident field--- also affects $Q_{\mathrm{s}}$.
 The effect of increasing $d_{Ti}$ is more significant when the field interacts from the Ti side. The solid red line shows $Q_{\mathrm{s}}$ for Al-Ti-Al trilayers.
At high thresholds $\nu_{g}\sim 70\,\,{\textrm{GHz}}$, corresponding to thin Ti layers,  $Q_{\mathrm{s}}$ for the trilayer lies between the bilayer $Q$-values. The
 increase of the trilayer $Q_{\mathrm{s}}$ compared to bilayers at low $\nu_{g}$  can be understood from Fig.~\ref{fig:Gap}: a thicker Ti layer is required in trilayers to achieve the same $\nu_{g}$. 
 Two points are important. $Q_{\mathrm{s}}$ for multilayer resonators with low $\nu_{g}$ remain highly-suitable for KIDs. The ordering of a \emph{bilayer} with respect to the field is also important to maximizing $Q_{\mathrm{s}}$. This emphasizes the need for full electromagnetic modeling of the field distributions.
\subsection{Quasiparticle Lifetimes in Multilayers}
\label{subsec:Life}
\begin{figure}[h]
\includegraphics[width=8.6cm]{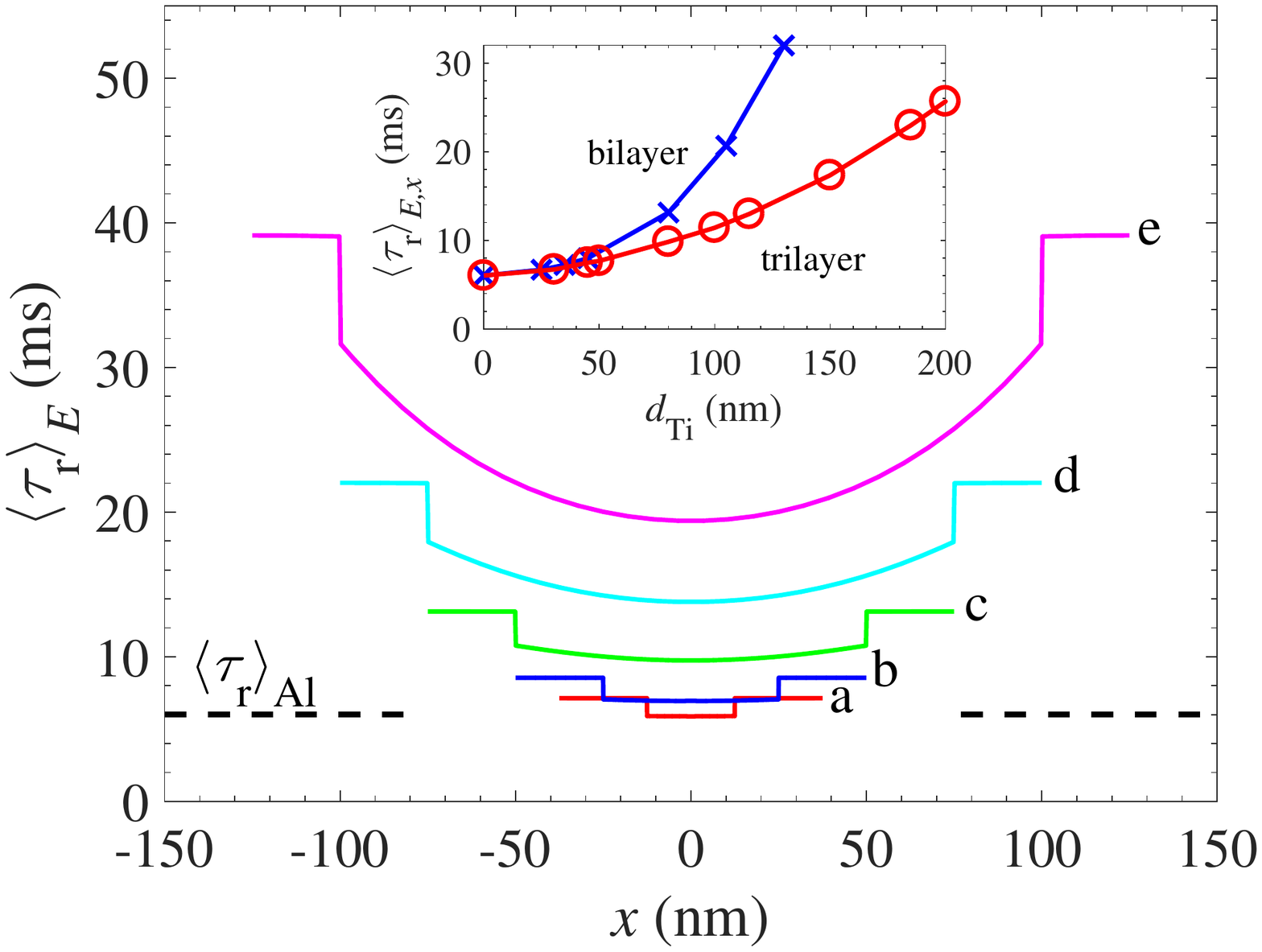}
\caption{\label{fig:Tau}
Energy-averaged recombination times $\langle \tau_{\mathrm{r}}\rangle_E$ for five proximity trilayers of $\mathrm{Al-Ti-Al}$ geometry, compared to that of thin-film Al (dashed line), as a function of position $x$.
(a) red line,~$d_{\mathrm{Ti}}=25\,\textrm{nm}$, (b) blue line,~$d_{\mathrm{Ti}}=50\,\textrm{nm}$, (c) green line,~$d_{\mathrm{Ti}}=100\,\textrm{nm}$, (d) cyan line,~$d_{\mathrm{Ti}}=150\,\textrm{nm}$, and (e) purple line,~$d_{\mathrm{Ti}}=200\,\mathrm{nm}$.
$\gamma_{\mathrm{B,Al}}=0.01$.
The temperature of the calculation is scaled such that $T = 170\times( \Delta_{\mathrm {g}}/\Delta_{\mathrm {Al}}) \,\,\textrm{mK}$ for each trilayer. The inset shows multilayer-averaged recombination times
$\langle \tau_{\mathrm{r}}\rangle_{E,x}$ as a function of Ti thickness for (i) red line, trilayer with total Al layer thickness $50\,\,\mathrm{nm}$, (ii) blue line, bilayer with Al layer thickness $50\,\mathrm{nm}$.
}
\end{figure}
\begin{figure}[h]
\includegraphics[width=8.6cm]{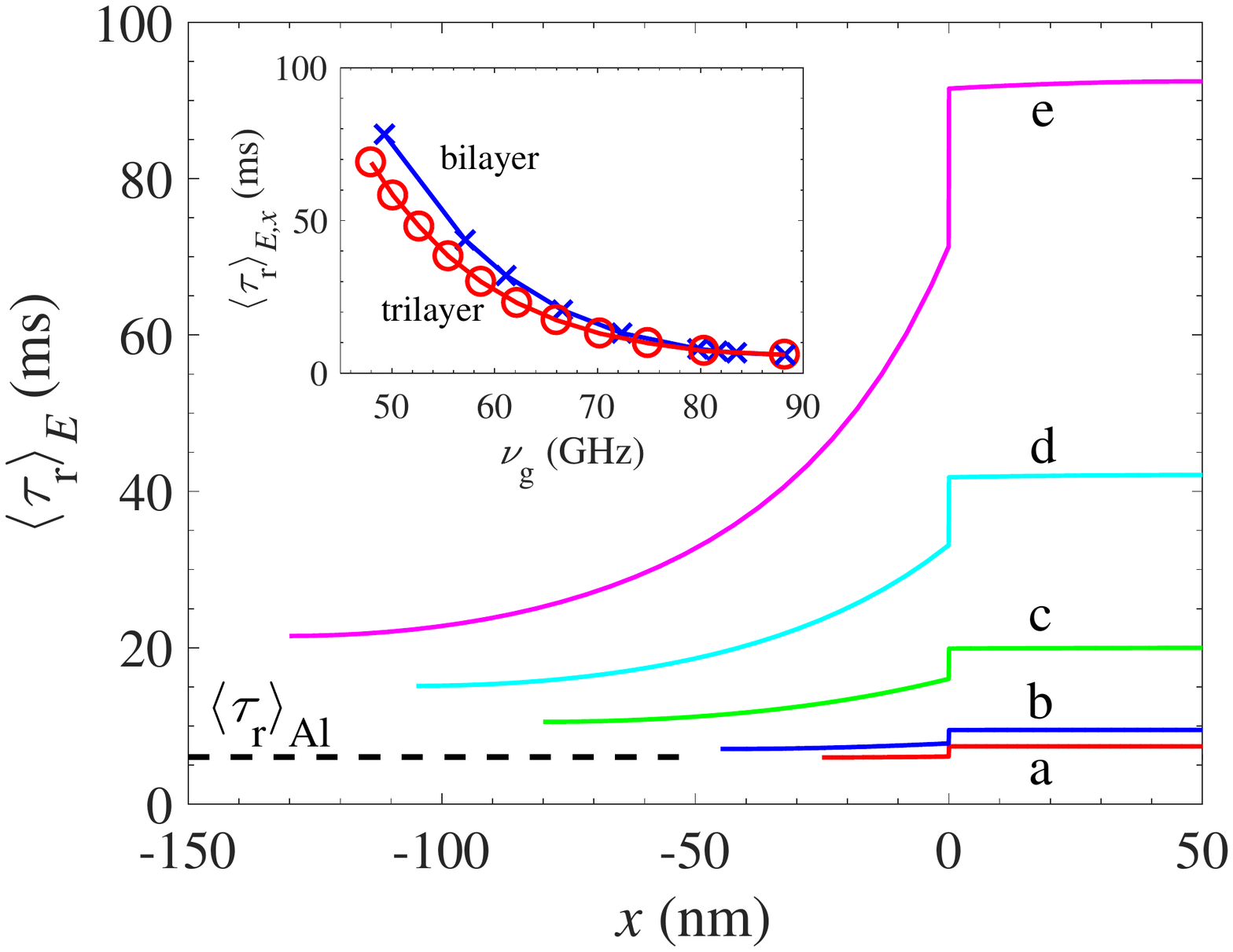}
\caption{\label{fig:Tau2}
Energy-averaged recombination times $\langle \tau_{\mathrm{r}}\rangle_E$ for five proximity bilayers of $\mathrm{Ti-Al}$ geometry, compared to that of thin-film Al (dashed line), as a function of position $x$.
(a) red line,~$d_{\mathrm{Ti}}=25\,\textrm{nm}$, (b) blue line,~$d_{\mathrm{Ti}}=45\,\textrm{nm}$, (c) green line,~$d_{\mathrm{Ti}}=80\,\textrm{nm}$, (d) cyan line,~$d_{\mathrm{Ti}}=105\,\textrm{nm}$, and (e) purple line,~$d_{\mathrm{Ti}}=130\,\textrm{nm}$.
$\gamma_{\mathrm{B,Al}}=0.01$.
The temperature of the calculation is scaled such that $T = 170\,\,\textrm{mK}\times (\Delta_{\mathrm {g}}/\Delta_{\mathrm {Al}})$ for each bilayer. The inset shows multilayer-averaged recombination times $\langle \tau_{\mathrm{r}}\rangle_{E,x}$ as a function of $\nu_{g}$ for (i) red line, trilayer with total Al layer thickness $50\,\mathrm{nm}$, (ii) blue line, bilayer with Al layer thickness $50\,\mathrm{nm}$.
}
\end{figure}
\noindent Here we show results of calculations of energy- and multilayer-averaged recombination  times $\langle\tau_{\mathrm{r}}(x)\rangle_E$  and $\langle \tau_{\mathrm{r}}\rangle_{E,x} $ respectively.
 For all multilayers,
 the temperature is scaled such that $T = 170\,\,\textrm{mK}\times (\Delta_{\mathrm {g}}/\Delta_{\mathrm {Al}})$.

 Figure~\ref{fig:Tau} shows  $\langle\tau_{\mathrm{r}}(x)\rangle_E$ as a function of position for \emph{trilayers} with
five Ti thicknesses
 (a) to (e) 25, 50, 100, 150 and $200\,\,\textrm{nm}$ respectively
 and, for comparison, calculation for Al.
For the thinnest trilayer, Fig.~\ref{fig:Tau}~(a) $\langle \tau_{\mathrm{r}}\rangle_E$ is close-to the calculated value for Al. For thicker trilayers $\langle \tau_{\mathrm{r}}\rangle_E$ \textit{increases} in both the Al and Ti layers. This arises from the reducing quasiparticle density of states in the Al at low energies as
 $d_{\mathrm{Ti}}$ increases
 (see Fig.~\ref{fig:DoS2_a}),
and the combined effect of the reduced lower limit of the integral in Eq.~(\ref{eq:Golubov_corrected}) and the $\Omega^2$ dependence of the phonon density of states factor within that integral.
 The situation is  similar for bilayers.

Figure~\ref{fig:Tau2} shows  $\langle\tau_{\mathrm{r}}(x)\rangle_E$ as a function of position for \emph{bilayers} for five 
 Ti thicknesses
 (a) to (e) 25, 45, 80, 105 and $130\,\,\textrm{nm}$ respectively and $d_{\mathrm{Al}}=50\,\,\textrm{nm}$.
The Ti thicknesses are chosen so that $\nu_{\mathrm{g}}$ for (a) to (e) in Fig.~\ref{fig:Tau2} are the same
as (a) to (e) in Fig.~\ref{fig:Tau}.
Although similar, in detail we see  that the bilayer recombination times are now  \emph{longer} in both the Ti and  Al compared to
 trilayers with the \emph{same} $\nu_{\mathrm{g}}$. 
This arises from the reduced DoS in the Al in bilayers compared to trilayers with the same $\nu_{\mathrm{g}}$,  noted earlier
in Sec.~\ref{subsec:DoS}.
This emphasizes the importance of a full calculation of the spatially-dependent DoS.

 The inset of Fig.~\ref{fig:Tau} shows $\langle \tau_{\mathrm{r}}\rangle_{E,x} $ as a function of $d_{\mathrm{Ti}}$ for bilayers and trilayers. As expected from the main plots,  $\langle \tau_{\mathrm{r}}\rangle_{E,x} $ increases with $d_{\mathrm{Ti}}$ and for bilayers   the  increase is more rapid.
The behaviour here can  be understood from Fig.~\ref{fig:Gap}, 
 $\Delta_{\mathrm{g}}$ changes more quickly with $d_{\mathrm{Ti}}$ for bilayers than trilayers.
The detailed differences arise from the different $d_{\mathrm{Ti}}$ required to achieve a given $\nu_{\mathrm{g}}$ in the two instances, the differing spatial variations in the DoS in Al and Ti, (in particular the reduced Al DoS in a bilayer having the same $\nu_{\mathrm{g}}$
  ---compare Figs.~\ref{fig:DoS2_a} and \ref{fig:DoS2-b}), and also the effect of the weighting with $N_0$ and $d_{\mathrm{Ti}}$ used in the calculation of $\langle \tau_{\mathrm{r}}\rangle_{E,x}$  in Eq.~(\ref{eq:tau_r_2}).

 The inset of Fig.~\ref{fig:Tau2} shows $\langle \tau_{\mathrm{r}}\rangle_{E,x}$ as a function of the experimental
  requirement $\nu_{g}$ for both bilayers and trilayers.
  We find that bilayers and trilayers with the same  $\nu_{g}$ have very similar multilayer-averaged recombination times, the bilayer being of order 20\% longer,
%
 %
so that both behave comparably. The choice between them can be made on the basis of other considerations - ease of deposition or material processing or film thickness requirements for example.
 We have not included the effect of phonon-trapping in these calculations.\cite{Rothwarf_Taylor} For thicker multilayers we would expect  phonon-trapping to provide an  additional enhancement of  the effective quasiparticle lifetime.

\section{Discussion and Conclusions}
\label{sec:conclusions}
\noindent We have described a full analysis of multilayer resonators suitable for KIDs based on the diffusive
 Usadel equations with appropriate boundary conditions that takes proper account of the spatial variation of the superconducting properties as a function position in the film.
We calculate the  spatial variation of the  superconducting order parameter $\Delta(x)$, the quasiparticle and pair densities of states, the superconducting energy gap $\Delta_{\mathrm{g}}$,
the complex conductivity $\sigma$, and energy- and multilayer-averaged $\tau_{\mathrm{r}}$ for Al-Ti multilayers. We  account for the spatial variation of
$T_{\textrm{c}}$, $N_0$, normal-state conductivity $\sigma_{\mathrm{N}}$, and characteristic quasiparticle lifetime
$\tau_0$.
We have also described how to calculate the surface impedance $Z_{\mathrm{s}}$ including   varying film properties
and the ordering of the multilayer with respect to the incident fields.
Our predictions of $\Delta_{\mathrm{g}}$ differ significantly from earlier predictions that were based on the weighted-average model, and indicate differences between bilayers and trilayers with the same total film thicknesses.

The calculated quasiparticle and pair densities of states for the Al-Ti multilayers considered here deviate significantly from the homogeneous (BCS) case.
We predict high quality factors $Q_{\mathrm{s}}$  
and long multilayer-averaged quasiparticle recombination times $\langle \tau_{\mathrm{r}}\rangle_{E,x}$ compared to thin-film Al.
Following  Ref. \citenumns{Leduc_Bumble_TiN}, we  define a  figure-of-merit for multilayers  such that  $\mathcal{F(\nu)} \propto Q_{\mathrm{s}}\langle \tau_{\mathrm{r}}\rangle_{E,x}\Theta(\nu-\nu_{\mathrm{g}})$, where $\Theta$ is the Heaviside step function,
giving appropriate weight to the required
detection threshold. Even without evaluation,  considering values of
$Q_{\mathrm{s}}$ and $\langle \tau_{\mathrm{r}}\rangle_{E,x}$ shown in Figs.~\ref{fig:Qs} and \ref{fig:Tau2},
values of $\mathcal{F}(\nu)$ for Al-Ti multilayers would be comparable to or better than those of Al films whilst achieving  the required
$\nu_{\mathrm{g}}$.
The multilayer structures described in this study are excellent  candidates for high-sensitivity,
 multiplexible  KIDs with targeted low-frequency detection thresholds.

Our work demands a programme of theoretical and experimental investigation of proximity-effect resonators for KID applications. 
 For the first time, full electromagnetic modeling of multilayer resonators is  possible.
We find that full consideration of the field distribution  is essential, particularly if  bilayers are to be used.
We  find that in terms of detector sensitivity, there should be little difference in the performance of bi- and trilayers
provided the geometry is properly considered.
 We have also  identified relatively simple experimental measurements 
 that would confirm our predictions of multilayer behaviour. Demonstrating high $Q$ and long $\tau_{\mathrm{r}}$ in Al-Ti multilayers is critical for KIDs based on  multilayers  to become the preferred solution for  sub-$100\,\,{\textrm{GHz}}$ detection. Further work is required to optimise device geometry and material combinations for particular applications.
Overall, we have described a coherent approach to calculating the  properties of superconducting multilayers for use as KIDs for experimentally important applications at frequencies in the range
$50-100\,\,\mathrm{GHz}$.
\bibliographystyle{h-physrev}
\bibliography{library_FINAL}

\begin{thebibliography}{10}

\bibitem{Jonas_review}
J.~Zmuidzinas,
\newblock {Annu. Rev. Condens. Matter Phys.} {\bf {3}}, 169 ({2012}).

\bibitem{Leduc_Bumble_TiN}
H.~G. Leduc {\em et~al.},
\newblock {Appl. Phys. Lett.} {\bf 97}, {102509} (2010).

\bibitem{Doyle_2008}
S.~Doyle, P.~Mauskopf, J.~Naylon, A.~Porch, and C.~Dunscombe,
\newblock J. Low Temp. Phys. {\bf 151}, 530 (2008).

\bibitem{Barends_tau_Q_2007}
R.~Barends {\em et~al.},
\newblock J. Low Temp. Phys. {\bf 151}, 518 (2008).

\bibitem{Janssen_2013}
R.~M.~J. Janssen {\em et~al.},
\newblock {Appl. Phys. Lett.} {\bf {103}} ({2013}).

\bibitem{deVisser_2014}
P.~J. de~Visser, J.~J.~A. Baselmans, J.~Bueno, N.~Llombart, and T.~M. Klapwijk,
\newblock {Nat. Commun.} {\bf {5}} ({2014}).

\bibitem{Planck_hifi_2011}
{Planck HiFi Core Team},
\newblock {Astron. Astrophys.} {\bf {536}}, {A4} ({2011}).

\bibitem{Catalano_2015}
A.~Catalano {\em et~al.},
\newblock {Astron. Astrophys.} {\bf {580}}, {A15} ({2015}).

\bibitem{Cicone_CO_2012}
C.~Cicone {\em et~al.},
\newblock {Astron. Astrophys.} {\bf {543}}, {A99} ({2012}).

\bibitem{Thomas_2014}
C.~N. Thomas {\em et~al.},
\newblock ({2014}), {astro-ph.IM/1401.4395v1}.

\bibitem{Mahfouf_2015}
J.~Mahfouf {\em et~al.},
\newblock {Q. J. R. Meteorol. Soc.} {\bf {141}}, 3268 ({2015}).

\bibitem{Aires_2015}
F.~Aires {\em et~al.},
\newblock {J. Geophys. Res. Atmos.} {\bf {120}}, 11334 ({2015}).

\bibitem{Turner_2016}
E.~C. Turner {\em et~al.},
\newblock {Atmos. Meas. Tech.} {\bf {9}}, 5461 ({2016}).

\bibitem{Tinkham_1994}
M.~Tinkham,
\newblock {\em Introduction to Superconductivity}, {2nd} ed. ({McGraw-Hill},
  New York, 1994).

\bibitem{Jones_2015}
G.~Jones {\em et~al.},
\newblock astro-ph.IM , 1701.08461v2 (2017).

\bibitem{Coiffard_2016_TiN}
G.~Coiffard {\em et~al.},
\newblock {J. Low Temp. Phys.} {\bf {184}}, 654 ({2016}).

\bibitem{Vissers_2013}
M.~R. Vissers {\em et~al.},
\newblock Thin Solid Films {\bf 548}, 485  (2013).

\bibitem{Vissers_highQ_Ti_TiN_2013}
M.~R. Vissers {\em et~al.},
\newblock Appl. Phys. Lett. {\bf 102} ({2013}).

\bibitem{Giachero_Ti_TiN}
A.~Giachero {\em et~al.},
\newblock J. Low Temp. Phys. {\bf 176}, 155 (2014).

\bibitem{Cooper_1961}
L.~N. Cooper,
\newblock {Phys. Rev. Lett.} {\bf {6}}, 68954 ({1961}).

\bibitem{Usadel1970}
K.~D. Usadel,
\newblock Phys. Rev. Lett. {\bf 25}, 507 (1970).

\bibitem{Brammertz2001}
G.~Brammertz {\em et~al.},
\newblock Physica C Supercond. {\bf 350}, 227 (2001).

\bibitem{Golubov2004}
A.~A. Golubov, M.~Y. Kupriyanov, and E.~Il'ichev,
\newblock Rev. Mod. Phys. {\bf 76}, 411 (2004).

\bibitem{Vasenko2008}
A.~S. Vasenko, A.~A. Golubov, M.~Y. Kupriyanov, and M.~Weides,
\newblock Phys. Rev. B {\bf 77}, 134507 (2008).

\bibitem{Wang_2017}
G.~Wang {\em et~al.},
\newblock {IEEE Trans. Appl. Supercon.} {\bf {27}} ({2017}).

\bibitem{George_2007}
G.~Vardulakis,
\newblock {\em Superconducting Kinetic Inductance Detectors: Theory,
  Simulations, and Experiments},
\newblock PhD thesis, University of Cambridge, 2007.

\bibitem{Monfardini_2011}
A.~Monfardini {\em et~al.},
\newblock {Astrophys. J. Suppl. S.} {\bf {194}} ({2011}).

\bibitem{Kuprianov1988}
M.~Y. Kuprianov and V.~F. Lukichev,
\newblock Sov. Phys. JETP {\bf 67}, 1163 (1988).

\bibitem{Nam1967}
S.~B. Nam,
\newblock {Phys. Rev.} {\bf {156}}, 487 ({1967}).

\bibitem{MattisBardeen}
D.~Mattis and J.~Bardeen,
\newblock {Phys. Rev.} {\bf {111}}, 412 ({1958}).

\bibitem{Kerr1996}
A.~R. Kerr,
\newblock National Radio Astronomy Observatory Report No. 245, 1996
  (unpublished).

\bibitem{KChang1994}
{Kai Chang},
\newblock {\em Microwave solid-state circuits and applications} ({Wiley}, New
  York, 1994).

\bibitem{Golubov1994}
A.~A. Golubov {\em et~al.},
\newblock {Phys. Rev. B} {\bf {49}}, 12953 ({1994}).

\bibitem{kaplan1976}
S.~B. Kaplan {\em et~al.},
\newblock Phys. Rev. B {\bf 14}, 4854 (1976).

\bibitem{Chang1978}
J.-J. Chang and D.~J. Scalapino,
\newblock J. Low Temp. Phys. {\bf 31}, 1 (1978).

\bibitem{Gladstone}
G.~Gladstone, M.~A. Jensen, and J.~R. Schrieffer,
\newblock Superconductivity in the transition metals: theory and measurement,
\newblock in {\em Superconductivity}, edited by R.~D. Parks, pp. 665--816,
  Marcel Dekker, New York, 1969.

\bibitem{Fujii2012}
G.~Fujii {\em et~al.},
\newblock {J. Low Temp. Phys.} {\bf {167}}, 815 ({2012}).

\bibitem{Parlato2005}
L.~Parlato {\em et~al.},
\newblock {Supercond. Sci. Technol.} {\bf {18}}, 1244 ({2005}).

\bibitem{Fukuda2007}
D.~Fukuda {\em et~al.},
\newblock {IEEE Trans. Appl. Supercon.} {\bf {17}}, 259 ({2007}).

\bibitem{Reale1973}
C.~Reale,
\newblock {Rev. Bras. Fis.} {\bf {3}}, 431 ({1973}).

\bibitem{de_Visser_2014}
P.~J. de~Visser {\em et~al.},
\newblock Phys. Rev. Lett. {\bf 112} (2014).

\bibitem{Rothwarf_Taylor}
A.~Rothwarf and B.~N. Taylor,
\newblock Phys. Rev. Lett. {\bf 19}, 27 (1967).

\end{thebibliography}
\end{document}